\documentclass[automark,headsepline,11pt,oneside,listsleft,DIV12]{scrartcl}
\usepackage{scrpage2}
\clearscrheadfoot
\ohead[]{\pagemark}
\usepackage{caption2}

\setcaptionwidth{1.0 \textwidth}

\usepackage{graphicx}
\usepackage{amssymb,amsmath} 
\usepackage{booktabs,units} 
\def\argmax{\ensuremath{\mathop{\text{arg}\,\text{max}}}}

\title{Estimating Acceleration and Lane-Changing Dynamics Based on NGSIM Trajectory Data}

\author{Christian Thiemann$^1$, Martin Treiber$^2$ and Arne Kesting$^3$\\[1ex]
$^1$Department for Nonlinear Dynamics\\
Max-Planck-Institute for Dynamics and Self-Organization\\
Bunsenstra{\ss}e 10, D-37073 G\"ottingen (Germany)\\[2ex]
$^{2,3}$Institute for Transport \& Economics\\
Technische Universit\"at Dresden (Germany)}
\date{March 30, 2008}
\begin{document}

\sloppy
\maketitle
\begin{abstract}
The NGSIM trajectory data sets provide longitudinal and lateral positional information for all vehicles in certain spatiotemporal regions. Velocity and acceleration information cannot be extracted directly since the noise in the NGSIM positional information is greatly increased by the necessary numerical differentiations. We propose a smoothing algorithm for positions, velocities and accelerations that can also be applied near the boundaries. The smoothing time interval is estimated based on velocity time series and the variance of the processed acceleration time series. The velocity information obtained in this way is then applied to calculate the density function of the two-dimensional distribution of velocity and  inverse distance, and the density of the distribution corresponding to the ``microscopic'' fundamental diagram. Furthermore, it is used to calculate the distributions of time gaps and times-to-collision, conditioned to several ranges of velocities and velocity differences. By simulating ``virtual stationary detectors'' we show that the probability for critical values of the times-to-collision is greatly underestimated when estimated from single-vehicle data of stationary detectors. Finally, we investigate the lane-changing process and formulate a quantitative criterion for the duration of lane changes that is based on the trajectory density in normalized coordinates. Remarkably, there is a very noisy but significant velocity advantage in favor of the targeted lane that decreases immediately before the change due to anticipatory accelerations.
\end{abstract}
\footnotetext[1]{E-mail: {\tt cthiema@nld.ds.mpg.de}}
\footnotetext[2]{E-mail: {\tt treiber@vwi.tu-dresden.de}, URL: {\tt http://www.traffic-simulation.de}}
\footnotetext[3]{E-mail: {\tt kesting@vwi.tu-dresden.de}, URL: {\tt http://www.akesting.de}}

%
\newpage
\section*{Introduction}
The Federal Highway Administration of the U.S.\ Department of Transportation
has originated the Next Generation SIMulation community (NGSIM) in order to
``improve the quality and performance of simulation tools, promote the use of
simulation for research and applications, and achieve wider acceptance of
validated simulation results''~\cite{NGSIM}.  As part of the program, a first
data set has been collected at the Berkeley Highway Laboratory (BHL) in
Emeryville by Cambridge Systematics and the California Center for Innovative
Transportation at UC Berkeley.  The BHL is a part of the I-80 at the east coast
of the San Francisco Bay.  Six cameras have been mounted on top of the
$\unit[97]{m}$ tall Pacific Park Plaza tower and recorded 4733 vehicles on a
road section of approximately $\unit[900]{m}$ length in a 30-minute period in
December~2003. The result has been published as the ``Prototype Dataset''.  As
part of the California Partners for Advanced Highways and Transit (PATH)
Program, the Institute of Transportation Studies at UC Berkeley further
enhanced the data collection procedure~\cite{skabardonis2005eav} and in
April~2005, another trajectory dataset was recorded at the same location using
seven cameras and capturing a total of 5648 vehicle trajectories in three
15-minute intervals on a road section of approximately $\unit[500]{m}$.  This
dataset was later published as the ``I-80 Dataset''.  In June~2005, another
data collection has been made using eight cameras on top of the $\unit[154]{m}$
tall 10~Universal City Plaza next to the Hollywood Freeway US-101.  On a road
section of $\unit[640]{m}$, 6101 vehicle trajectories have been recorded in
three consecutive 15-minute intervals. This dataset has been published as the
``US-101 Dataset''.  All datasets are freely available for download at the
NGSIM homepage (\texttt{www.ngsim.fhwa.dot.gov}).

This amount of trajectory data is so far unique in the history of traffic
research and provides a great and valuable basis for the validation and
calibration of microscopic traffic models and already received some amount of
attention.  For example, Lu and Skabardonis examined the backward propagation
speed of traffic shockwaves using the two later datasets~\cite{lu2007fts}.
However, most recent attention focuses on the investigation of lane changes:
Roess and Ulerio have used the two later datasets to study some trends and
sensitivities in weaving sections~\cite{roess2007aof}, especially lane changes.
Zhang and Kovvali~\cite{zhang2007fga} and Goswami and
Bham~\cite{goswami2007gab} investigated the gap acceptance behavior in
lane-changing situation on freeways. Using the Prototype and I-80 datasets,
Toledo and Zohar investigated the duration of lane
changes~\cite{toledo2007mdo}.  Choudhury~et~al.\ have calibrated a lane
changing model using the I-80 dataset and validated the model using virtual
loop detectors placed into the US-101 data~\cite{choudhury2007mcl}.
Leclercq~et~al.~\cite{leclercq2007rpa} have calibrated a model of the headway
relaxation phenomenon observed in lane-changing situations using the I-80
dataset.  Further studies using the NGSIM data include
Refs.~\cite{vu2007sow,jin2007sof,webster2007tdl,alecsandru2007ard}.

In all of the above work, the longitudinal and lateral position
information of the trajectory data has been used essentially
directly. In contrast, there are very few investigations of the data
with respect to topics where velocities and accelerations play a
significant role such as testing or calibrating car-following models
\cite{Kesting-Calibration-TRB08} or lane-changing models, or estimating
fuel consumption \cite{Treiber-Fuel-TRB08}. Since velocities and
accelerations are derived quantities, the noise in the NGSIM
positional information is greatly increased and a direct application
is not possible. 

In this work, we will first propose and motivate a smoothing method that
enables the NGSIM data to be used for data analysis using the velocity
or acceleration information. The smoothed velocities will then be used to calculate
the density function of the two-dimensional distribution of velocity
and inverse distance, and the density of the distribution
corresponding to a ``microscopic'' fundamental diagram.  The smoothed data will
also be used to calculate the distributions of time gaps and
times-to-collision, conditioned to several ranges of velocities and
velocity differences. Furthermore, we will compare the measurements of
spatial quantities by virtual loop detectors with their real values
determined from the trajectory data. Finally, we will propose a method
to determine the lane change duration from the NGSIM data. We will
close with a discussion of the findings and suggestions for future
research problems.

\section*{Extracting the Velocity and Acceleration Information}

\begin{figure}[t]
  \centering
  \includegraphics[width=.475\linewidth]{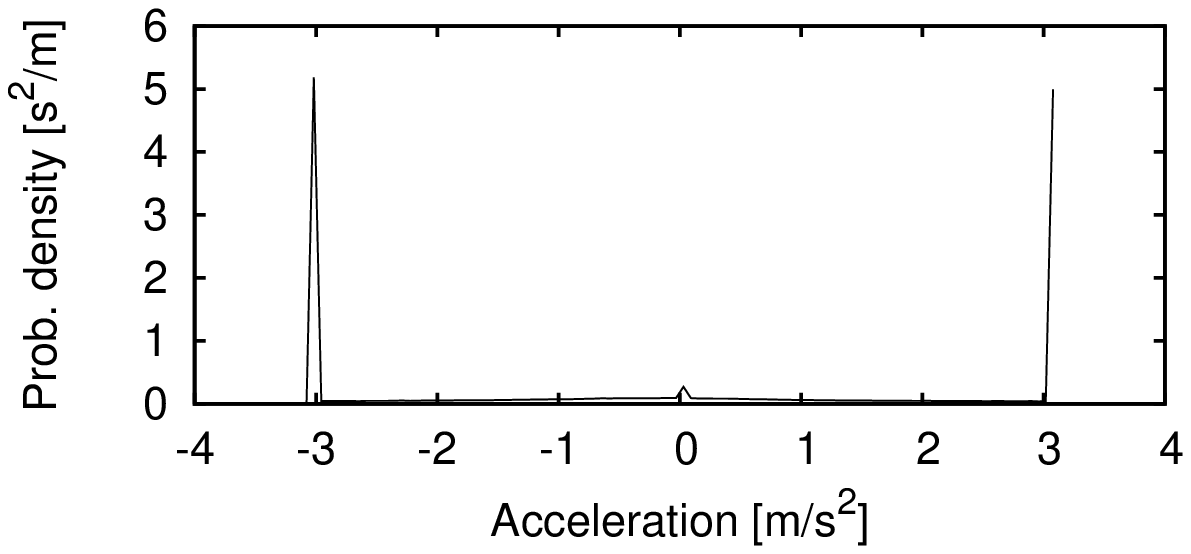}\hfill
  \includegraphics[width=.475\linewidth]{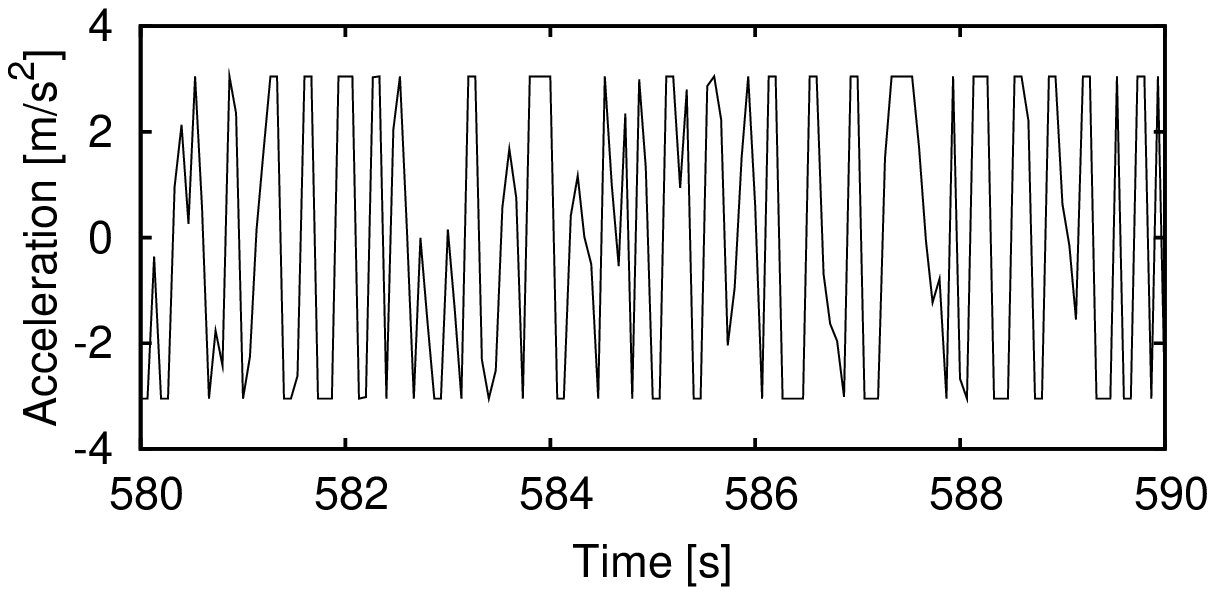}
  \includegraphics[width=.475\linewidth]{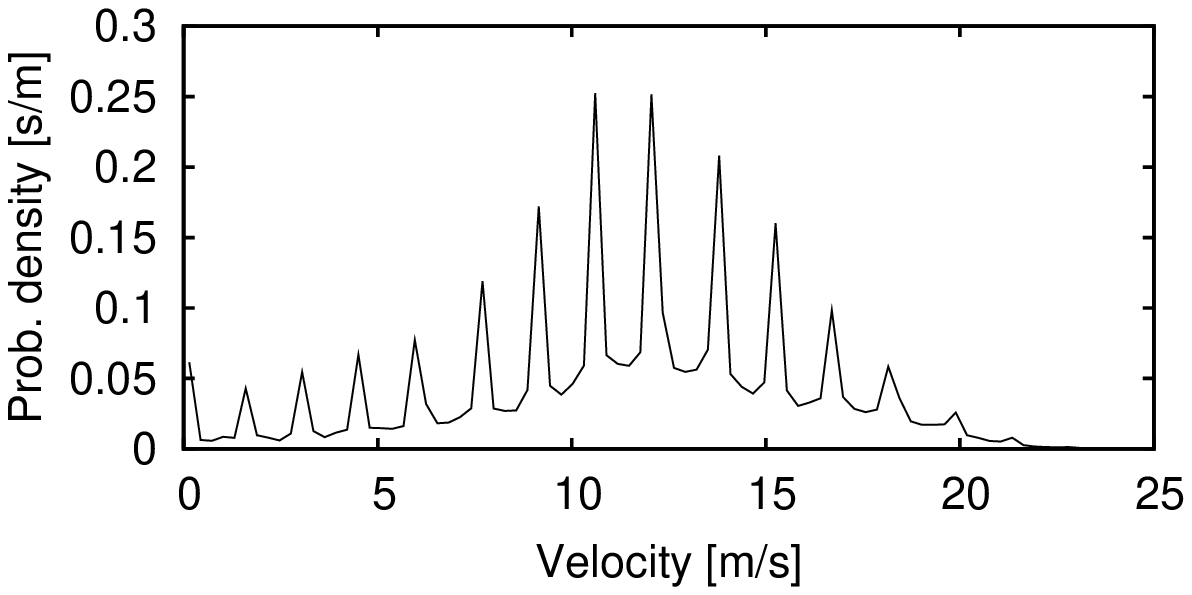}\hfill
  \includegraphics[width=.475\linewidth]{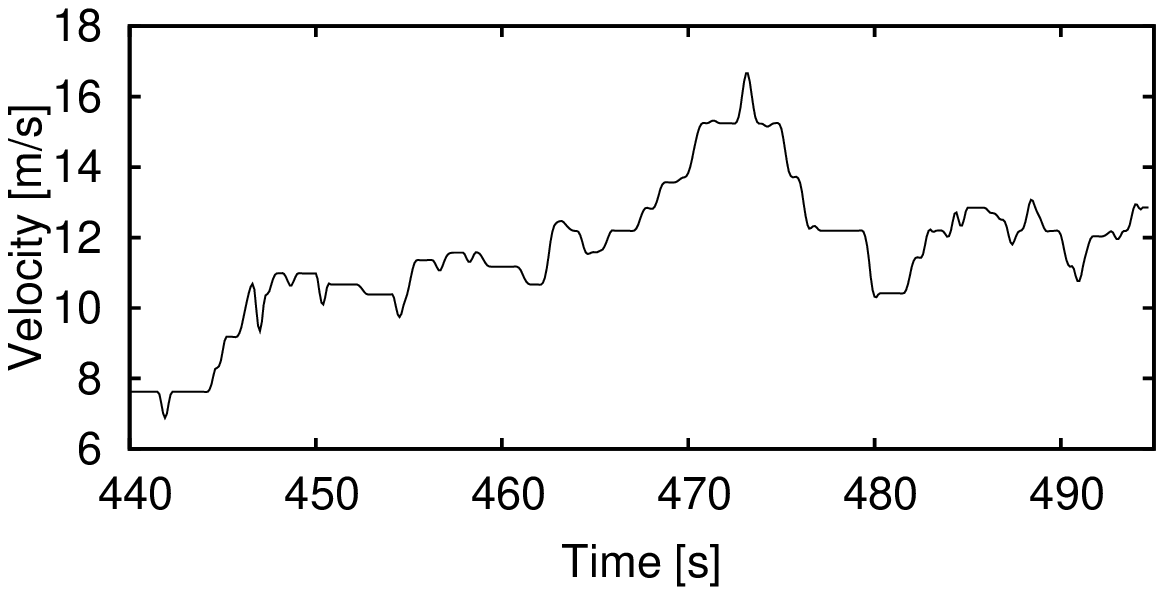}
  \caption{\label{fig:originaldata}Problems of the original, unsmoothed data:
  In the top row we visualize the unrealistic acceleration found in the
  Prototype dataset.  The top left plot shows the acceleration distribution,
  the top right shows an example trajectory excerpt.  In the bottom row we
  show the velocity distribution of the 7.50am--8.05am datafile of the US-101
  dataset on the left and an example trajectory (velocity) on the right.}
\end{figure}

The trajectory data available for download seems to be unfiltered and exhibits
some noise artefacts. All data sets include velocity and acceleration.
However, they seem to have been numerically derived from the tracked vehicle
positions without any processing.  Fig.~\ref{fig:originaldata} visualizes the
problems of the data: In the Prototype dataset two thirds of all accelerations
are beyond $\unit[\pm 3]{m/s^2}$ (which are then reported as $\unit[\pm
3]{m/s^2}$ in the datafile), as can be seen from the acceleration
distribution.  The example trajectory shows that the driver is allegedly
changing between hard acceleration and hard deceleration several times a
second which is clearly unrealistic.  In the later I-80 and US-101 datasets,
the acceleration distributions are more realistic -- though approximately 10\%
are beyond $\unit[\pm 3]{m/s^2}$.  However, in the later datasets, the
velocity distributions are very spiky, i.e., velocities tend to snap to
certain values.  Looking at the velocities of an example trajectory exhibits
an unrealistic behavior: If taken for real, this would mean that drivers do
not smoothly brake or accelerate but use the gas and brake pedal only
occasionally but hard to quickly change between ``preferred velocities''.
Also, to produce the spikes in the velocity distribution, all drivers must
happen to ``like'' the same velocities.  This is clearly unrealistic and
therefore the velocity spikes must be an artefact of the measurement method.
One may credit the velocity spikes to discretization errors (time and space
are discretized, thus velocity can only take certain discrete values as well),
but two observations object to that: First, the spikes are not delta peaks,
other velocities still do appear.  Second, given the time discretization $dt =
\unit[\frac{1}{10}]{s}$ and the approximate distance between the velocity
spikes $dv \approx \unit[0.7]{m/s}$, this would mean that the spatial accuracy
of the measurement method is just $\unit[7]{m}$ (which is obviously not the
case).  We therefore suspect that the velocity spikes are introduced by some
data post-processing.

In order to correct those artefacts, we have applied a {\it symmetric
exponential moving average filter (sEMA)} to all trajectories before any
further data analysis.  This process is presented in this section.

Let $x_\alpha(t_i)$ denote the measured position of vehicle $\alpha$
at time $t_i$, where $i = 1\dots N_\alpha$ and $N_\alpha$ denotes the
number of datapoints of the trajectory.  The smoothing kernel is given
by $g(t) = \exp(-|t|/T)$ where $T$ is the smoothing width.  Since the
datapoints are equidistant in time with interval $dt$, we can
formulate the smoothing operation by using datapoint indices instead
of times. The smoothed positions $\tilde{x}(t_i)$ are given by
\begin{equation}\label{eq:smoothing}
    \tilde x_\alpha(t_i) = \frac{1}{Z}\,\sum_{k = i-D}^{i+D} x_\alpha(t_k)\,e^{-|i-k|/\Delta}
    \quad\text{where } Z = \sum_{k = i-D}^{i+D} e^{-|i-k|/\Delta}.
\end{equation}
The smoothing width $\Delta$ is given by $T/dt$ and transparently handles the
different time intervals in the datasets (the Prototype dataset uses $dt =
\unit[\frac{1}{15}]{s}$ while the later two use $dt =
\unit[\frac{1}{10}]{s}$): We can use the same real time smoothing width $T$
for all datasets and $\Delta = T/dt$ will be the corresponding smoothing width
measured in datapoints for the specific dataset.  The smoothing window width
$D= \max\{ 3\,\Delta, i-1, N_\alpha-i \}$ is chosen to be three times the
smoothing kernel width for any data point that is not closer than $D$ data
points to either trajectory boundary. For the points near the boundaries, the
smoothing width is decreased to ensure that the smoothing window is always
symmetric. 

It may be objected that other filters would work as well or even better, e.g.,
the Kalman filter or a simple moving average.  A moving average filter, which
would correspond to Eq.~\eqref{eq:smoothing} with the exponentials removed, has
non-continuous filter boundaries, i.e., with moving the filter data points
suddenly slip into the smoothing window with full weight or suddenly drop out.
This can cause smoothing artefacts which are prevented by using a weighted
moving average where the weight decreases with increasing distance from the
smoothing window center.  This way data points will be smoothly incorporated
into the smoothing window and fall out smoothly as well.  We found that an
exponential weight function leads to better results than a gaussian filter,
thus we decided for the sEMA.  The Kalman filter needs a simple traffic model
and thus introduces some significant assumptions into the smoothing process.
Also, the Kalman filter has more parameters while the sEMA method has only one
parameter, $T$, and does not introduce complicated assumptions.

Another possible filter would be to not use some moving kernel filter but
increase the step size from $dt$ to $n\,dt$ in calculation of the velocities
and accelerations, i.e., $v(t) = (x(t+n\,dt)-x(t-n\,dt))/(2\,n\,dt)$.  It can
be shown that this filter is equal to a simple moving average for the
velocities and a composition of two moving averages for the accelerations
(which simplifies to a triangular moving average when boundary regions are
neglected).  This filter is a faster but somewhat worse alternative to our
proposed method.

Having defined the fundamental smoothing mechanism, there are still two open
questions: First, the order of differentiations and smoothing operations need
to be defined, and second, a smoothing width $T$ must be found.

Addressing the first question, there are three possible answers: (i)
Smooth positions, then differentiate to velocities and accelerations,
(ii) first differentiate to velocities and accelerations and then
smooth all three variables, or (iii) smooth positions, differentiate
to velocities, smooth velocities, differentiate to accelerations and
smooth accelerations. For $D+2\le i \le N_\alpha-D-1$, the
smoothing~\eqref{eq:smoothing} commutes with the differentiation, and
all these methods are equivalent. In view of the short trajectories,
however, the points closer to the boundary cannot be neglected.

The first method is very problematic as can be seen by the following
reasoning. Consider an artificial trajectory with constant
acceleration: $x(t_i) = \frac{1}{2}\,a\,t_i^2$.  Any symmetric
smoothing kernel will overestimate the position and produce a
trajectory $\tilde x(t_i) > x(t_i)$. Sufficiently far away from the
boundaries, the smoothing window width $D$ is constant and the
smoothed trajectory $\tilde{x}(t_i) = x(t_i) +\frac{1}{2} a
\sigma_g^2$ has a constant error proportional to the variance
$\sigma_g^2$ of the smoothing kernel. Near the boundaries, however,
$D$ and $\sigma_g$ will become smaller and vanish for $i = 1$ and $i =
N$, which results in $\tilde x(t_1) = x(t_1)$ and $\tilde x(t_N) =
x(t_N)$.  Thus, the offset of the smoothed positions becomes smaller
when approaching the boundaries, which of course induces a bias to the
velocity.  Moreover, if the smoothing kernel does not completely
vanish at the smoothing window borders, the transition between
constant offset and decreasing offset will not be continuously
differentiable inducing a jump in velocity and thus an even larger
jump in the acceleration. Therefore, we discourage from this smoothing
method.

In order to decide for the second or third smoothing method, we have
generated artificial benchmark trajectories and added some white noise to the
positions.  The second method -- first the differentiation to
velocities and accelerations and then the smoothing of the three
variables -- turned out to better reproduce the original trajectories,
thus we decided to use this method.

\begin{figure}
  \centering
  \includegraphics[width=.475\linewidth]{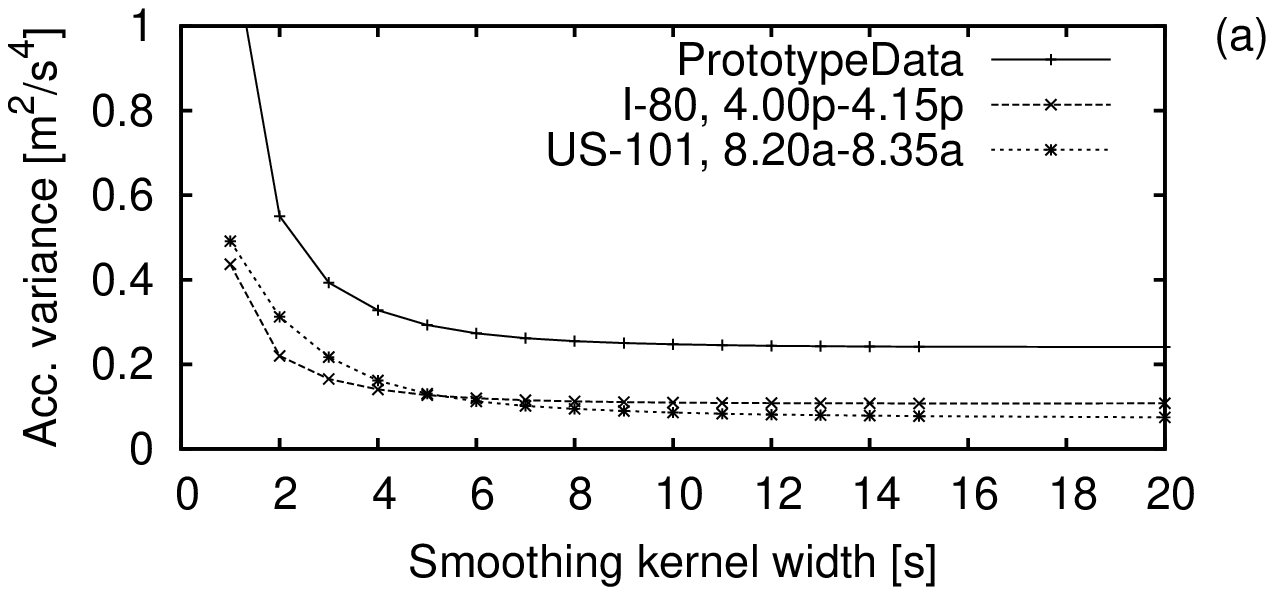}\hfill
  \includegraphics[width=.475\linewidth]{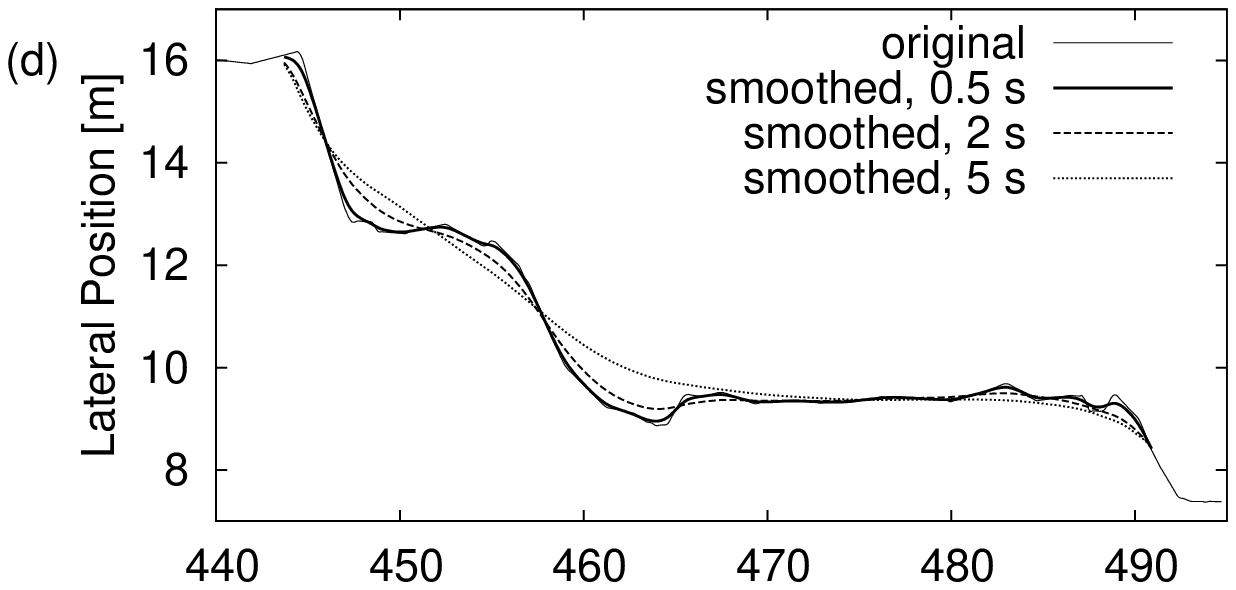}\\[-2ex]
  \includegraphics[width=.475\linewidth]{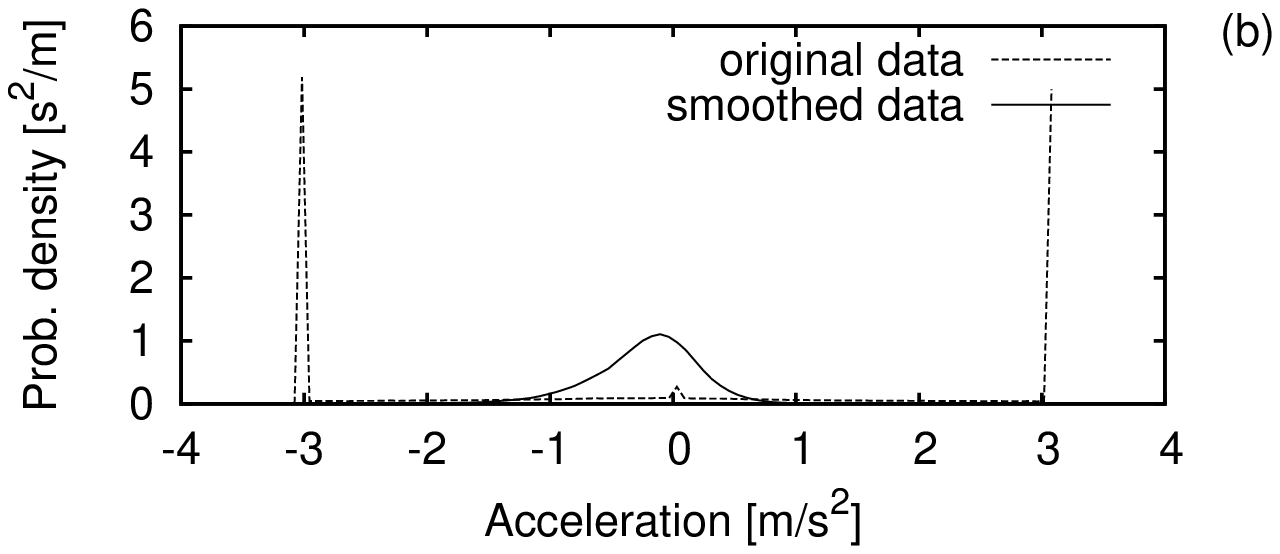}\hfill
  \includegraphics[width=.475\linewidth]{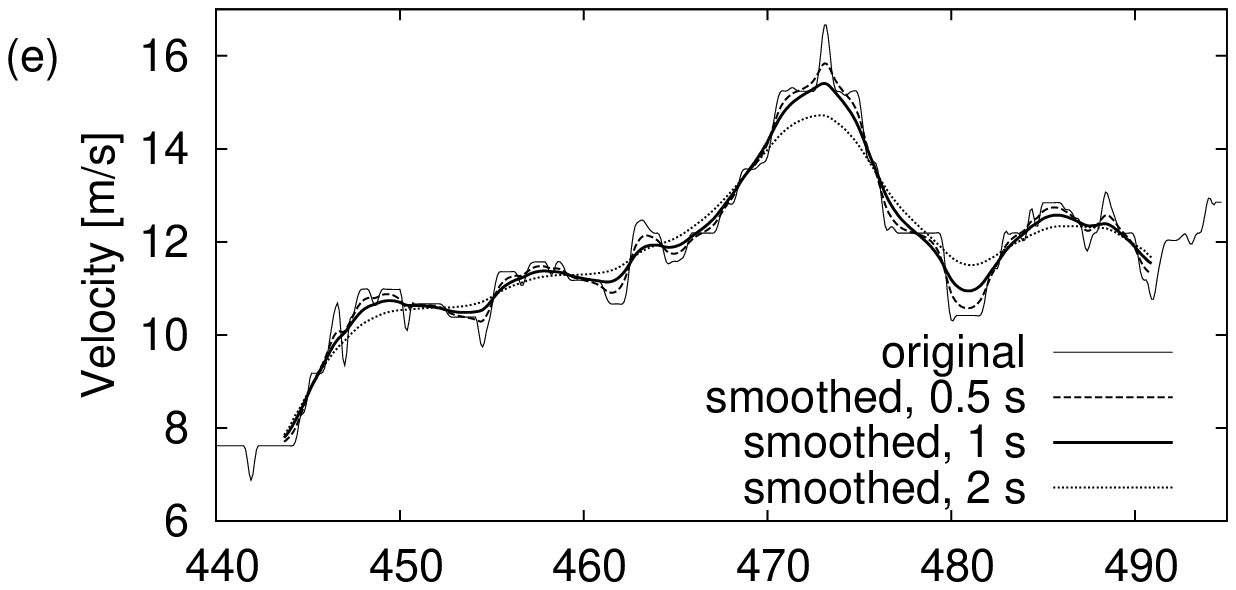}\\[-2ex]
  \includegraphics[width=.475\linewidth]{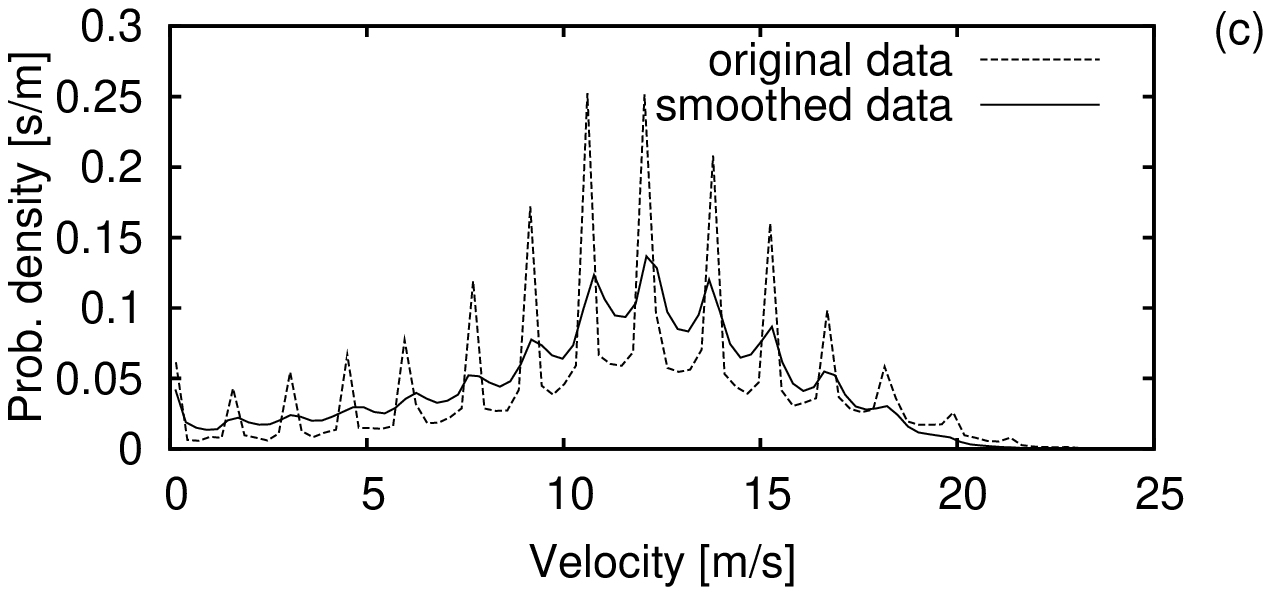}\hfill
  \includegraphics[width=.475\linewidth]{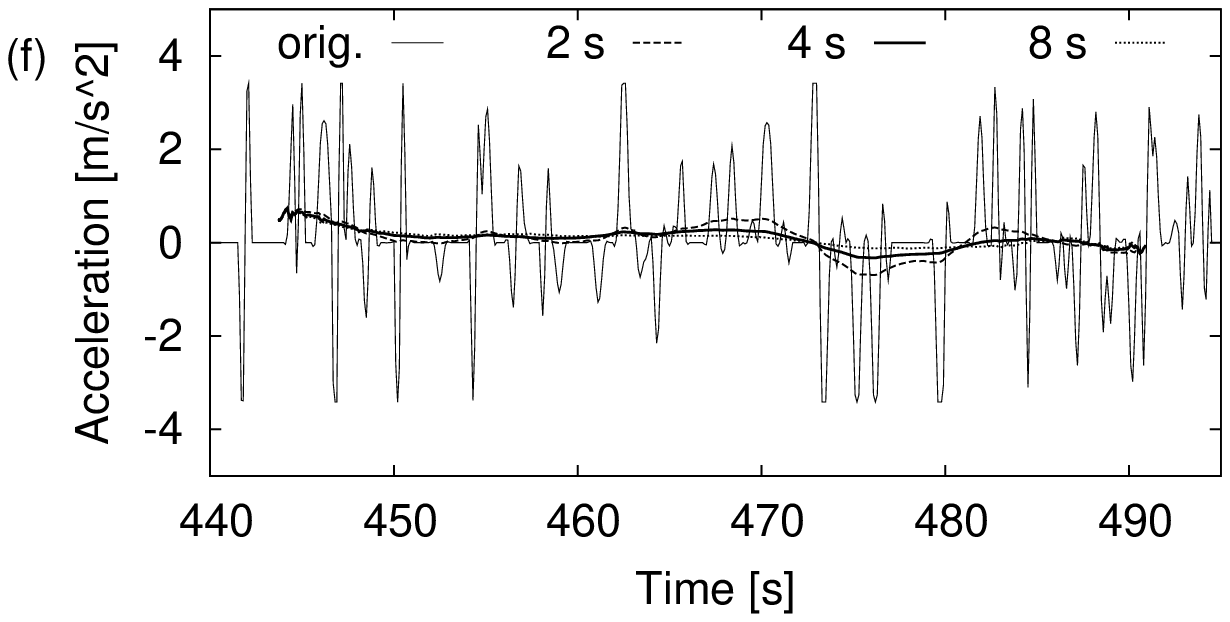}

  \caption{\label{fig:smoothing}Effects of the applied trajectory smoothing:
  (a) The dependency of the acceleration variances on the smoothing kernel
  width.  (b) Acceleration distribution in the Prototype dataset and (c)
  velocity distribution in the 7.50am--8.05am datafile of the US-101 dataset.
  In the right column the lateral position (d), longitudinal velocity (e) and
  acceleration (f) of a sample trajectory of the US-101 dataset with different
  applied smoothing kernel widths is shown.}
\end{figure}

This left us with the difficult question of which smoothing width $T$
to use.  There is no generic recipe, but we collected some hints that
helped making this decision not completely arbitrary.  First, we
extracted the most ``vivid'' trajectories -- those with a large
velocity range -- from each dataset and compared the variance of the
accelerations, $\sigma^2_a$, for different smoothing widths
(cf.~Fig.~\ref{fig:smoothing}(a)).  For $T\to\infty$, the
acceleration variance of the smoothed trajectory would vanish, but the
variance that is caused by the noise vanishes much faster than the one caused
by the real acceleration data.  Thus, with finite $T$ the noise is smoothed out
very quickly, leading to a fast drop in $\sigma^2_a(T)$ at small $T$.
For larger $T$, $\sigma^2_a(T)$ appears to be nearly constant.
Keeping in mind that the real acceleration data is smoothed a little bit as well, the
plot suggests a smoothing width of about $\unit[4]{s}$.

However, this value is a suggestion for the {\it acceleration}
smoothing width only.  We will now show that it is not necessary to
use such large smoothing widths for the positions and velocities.  Let
$X_\alpha(t_i)$ be a random variable describing the positions of
vehicle $\alpha$ with expectation value $\bar x_\alpha(t_i)$ and variance
$\sigma^2_x(t_i)$.  The measured trajectory $x_\alpha(t_i)$ is a
realization of $X_\alpha(t_i)$ and, assuming unbiased noise, the real
trajectory is equal to $\bar x_\alpha(t_i)$.  Now, we define two new
random variables describing the velocities and accelerations of
vehicle $\alpha$ in terms of symmetric difference quotients,
\begin{align}
  V_\alpha(t_i) &= \frac{X_\alpha(t_i+dt) - X_\alpha(t_i-dt)}{2\,dt}, \\
  A_\alpha(t_i) &= \frac{X_\alpha(t_i+dt) - 2\,X_\alpha(t_i) + X_\alpha(t_i-dt)}{dt^2}.
\end{align}
Since this is a linear combination of random variables, the
expectation values of $V_\alpha(t_i)$ and $A_\alpha(t_i)$ will be the
first and second derivative of $\bar x_\alpha(t_i)$, respectively (the
real velocities and the real accelerations). Assuming uncorrelated
noise, the variances of $V_\alpha(t_i)$ and $A_\alpha(t_i)$ are given
by
\begin{equation}
  \sigma^2_V(t_i) = \frac{\sigma^2_x(t_i)}{2\,dt^2}\quad\text{and}\quad
  \sigma^2_A(t_i) = \frac{6\,\sigma^2_x(t_i)}{dt^4}.
\end{equation}
Thus, the noise will be strongly amplified by the differentiation and
therefore, the velocities must be weaker smoothed than the
acceleration and the positions weaker than the
velocities. 

In Fig.~\ref{fig:smoothing}(b,c) we plotted the
lateral positions and longitudinal velocities and accelerations of a
sample trajectory of the US-101 dataset as original data and for different
smoothing widths.  The position smoothing width $T_x$ is very
critical, because the lane change duration is quite sensitive to it.
As visible in the plot, a large $T_x$ will significantly smear out the
trajectory leading to larger lane change durations.  In order to resolve the
issue with the ``preferred velocities'' the
smoothing of the velocities should be strong enough so that the
smoothed velocities no longer follow the trends of this
``semi-quantization''. However, the smoothing should be as weak as
possible because the velocity smoothing width also quantitatively
influences some results.  Finally, we decided in favor of the
smoothing times
\begin{equation}
  T_x = \unit[0.5]{s}\text{, }\,
  T_v = \unit[1]{s}\text{,}\,\text{ and }\,
  T_a = \unit[4]{s}.
\end{equation}
The effects of this smoothing on the acceleration distribution of the
Prototype dataset and the velocity distribution of the two later datasets can
be seen in Fig.~\ref{fig:smoothing}(d-f).

\clearpage
\section*{Results}

Most empirical traffic state data is gathered by stationary loop
detectors that can measure quantities at different times, but at a
single location only.  These measurement devices are therefore capable
of measuring temporal quantities, but not spatial quantities.
However, since both spatial and temporal quantities are important in
traffic science, it is common practice to derive the spatial
quantities from temporal measurements by using some conservation
assumptions (e.g., constant vehicle velocities within a certain time
period).  Modern trajectory data like the NGSIM recordings provide
enough data to enable a validation of these practices.

In the following, we will describe the analysis process to obtain
spatial information from temporal data, and vice versa, and check its
accuracy for three examples: The microscopic fundamental diagram, and
the distributions of the time gaps and times-to-collision.  Later, we
will investigate lane changes in the NGSIM data.  All following analysis will
use the smoothed datasets obtained by the smoothing method introduced and
motivated above---and all references to any ``NGSIM dataset'' are to be
understood as references to the smoothed datasets.

\subsection*{Spatial and Temporal Quantities from Momentary and Stationary Measurements}

\begin{figure}
  \centering
\begin{tabular}{cc}
  \includegraphics[width=0.475\linewidth]{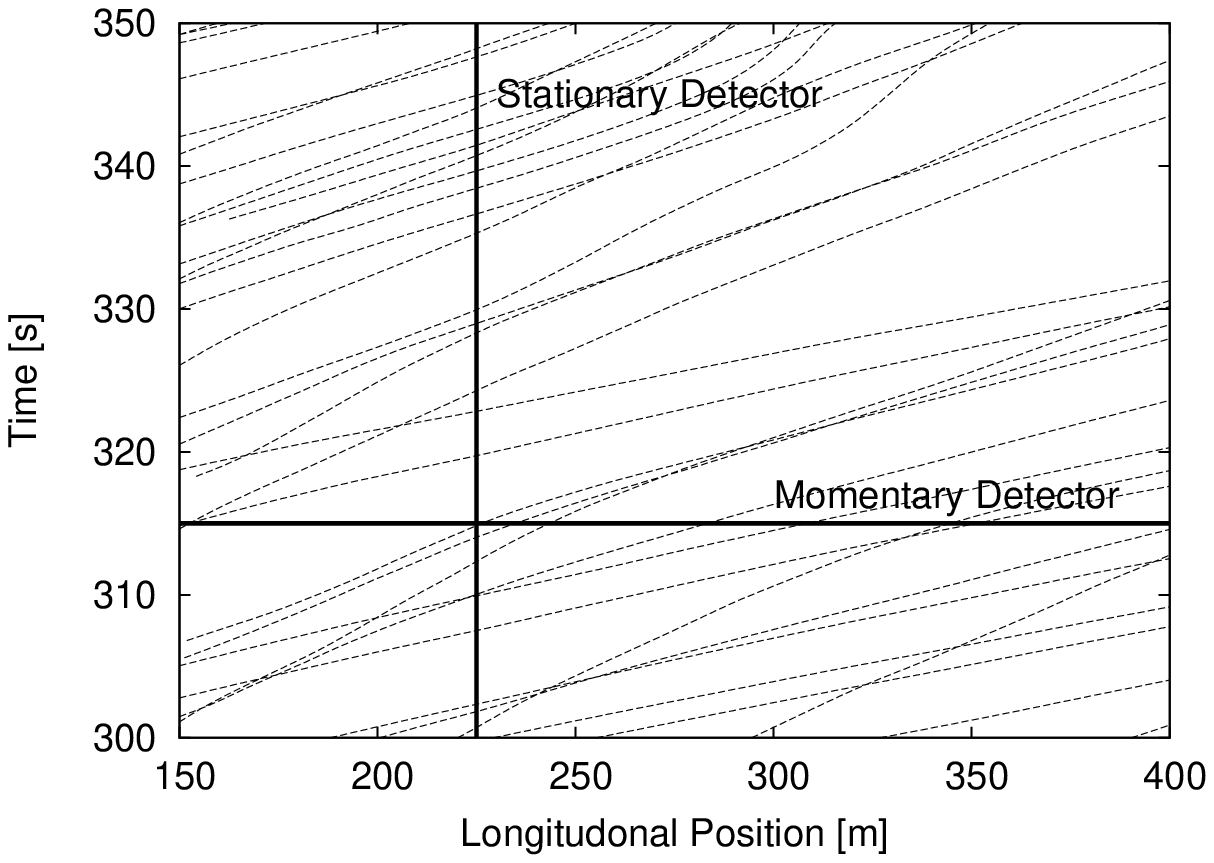} &
  \includegraphics[width=0.475\linewidth]{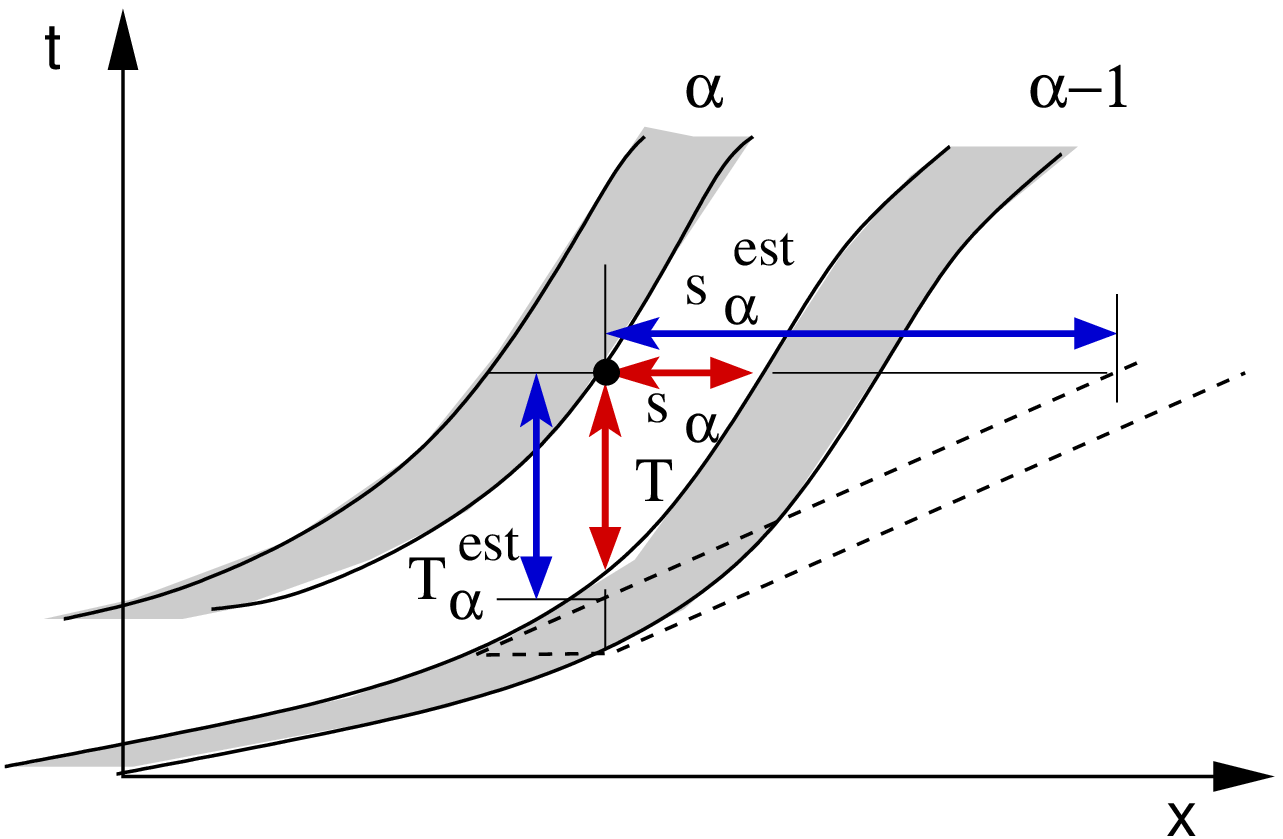}
\end{tabular}
  \caption{\label{fig:virtualdetectors}Left: Virtual detectors in the
  space-time-plot: Stationary detectors (loop detectors) correspond to
  lines parallel to the time axis while momentary detectors (aerial
  photographs) correspond to lines parallel to the space axis. Right:
  Illustration of the time gap $T^\text{est,pt}_\alpha$ according to
  Eq.~\eqref{eq:timegap_pt} assuming constant velocities and the real
  time gap $T$ (please note that $T^\text{est,pt}_\alpha$ is an estimate
  from \emph{stationary measurement} while $T^\text{est,mom}_\alpha$ as
  defined in Eq.~\eqref{eq:timegap_mom} is an estimate from momentary
  measurement).}
 
\end{figure}

The two measurement types we want to compare are the traditional
stationary loop detector, which is singular in space but continuous in
time, and an aerial photograph, which is continuous in space but
singular in time.  The basic idea of our analysis is to place virtual
loop detectors into the trajectory data.  These would correspond to
lines parallel to the time axis in a space-time-plot, while lines
parallel to the space axis correspond to momentary snapshots (virtual
photographs) of the measurement area
(cf.~Fig.~\ref{fig:virtualdetectors}).  Wherever those lines
intersect, both stationary and momentary measurements are available
for comparison.  To maximize the amount of data available for
comparison, we applied the following algorithm to the data: For every
tenth datapoint of each trajectory, the spatial leader and the
temporal leader are determined.  The spatial leader $\alpha-1$ is the
vehicle currently driving ahead of the vehicle $\alpha$ and the
temporal leader is the vehicle that most recently passed the actual
position of vehicle $\alpha$ (for simplicity, we will denote the
temporal leader with $\alpha-1$ as well).  The first information is
only available to momentary measurements while the second is only
available to stationary measurements.

Assuming double loop detectors for the {\it stationary} measurement,
the passage times $t_\alpha$ and $t_{\alpha-1}$ of vehicle $\alpha$
and $\alpha-1$ and their velocities at the time of passing the
detector are available: $v_\alpha(t_\alpha)$,
$v_{\alpha-1}(t_{\alpha-1})$.  Furthermore, we know the length of the
leading vehicle $l_{\alpha-1}$ and, of course, the positions (front
bumper) at the time of passing the detector: $x_\alpha(t_\alpha) =
x_{\alpha-1}(t_{\alpha-1})$.

From the momentary measurement at time $t_\alpha$ we obtain the
positions of the two vehicles, $x_\alpha(t_\alpha)$ and
$x_{\alpha-1}(t_\alpha)$, as well as the length of the leading vehicle
$l_{\alpha-1}$.  Assuming that we take two consecutive photographs, we
can also determine the velocities $v_\alpha(t_\alpha)$ and
$v_{\alpha-1}(t_\alpha)$.  From this momentary measurement, the
following {\it spatial} quantities can be calculated:
\begin{align}
 \label{eq:spat_real_s}
 \text{Spatial gap }\quad s_\alpha(t_\alpha) &= x_{\alpha-1}(t_\alpha) - x_\alpha(t_\alpha) - l_{\alpha-1}, \\
 \label{eq:spat_real_v}
 \text{Approaching rate } \Delta v_\alpha(t_\alpha) &= v_\alpha(t_\alpha) - v_{\alpha-1}(t_\alpha). 
\end{align}

Assuming constant velocities within the time interval $\Delta t_\alpha
= t_\alpha - t_{\alpha-1}$, we can estimate the same quantities from
data collected by a stationary detector:
\begin{align}
 \label{eq:spat_est_s}
 s^\text{est}_\alpha(t_\alpha) &= v_{\alpha-1}(t_{\alpha-1})\,\Delta t_\alpha - l_{\alpha-1},\\
 \label{eq:spat_est_v}
 \Delta v^\text{est}_\alpha(t_\alpha) &= v_\alpha(t_\alpha) - v_{\alpha-1}(t_{\alpha-1}).
\end{align}

Furthermore, the time gap $T$ defined by the gap related to the actual
velocity, $s/v$, is a crucial quantity for the safety and capacity of
traffic flow. From the time interval between two vehicles passing the
stationary detector, $\Delta t_\alpha=t_\alpha - t_{\alpha-1}$, we can
estimate the time gap while passing the detector:
\begin{equation}
 T^\text{est,pt}_\alpha(t_\alpha) = \Delta t_\alpha - \frac{l_{\alpha-1}}{v_{\alpha-1}(t_{\alpha-1})}.
 \label{eq:timegap_pt}
\end{equation}
This definition assumes constant velocity of the leading vehicle in
the time interval $\Delta t_\alpha$. The ``real'' time
gap, however, would be obtained by measuring the time where the rear
bumper of the leading vehicle passed the detector:
\begin{equation}\label{eq:timegap}
 T_\alpha(t_\alpha) = t_\alpha - t' \quad\text{with $t'$ such
 that}\quad x_{\alpha-1}(t') - l_{\alpha-1} = x_\alpha(t_\alpha).
\end{equation}
Both quantities are illustrated in Fig.~\ref{fig:virtualdetectors}.
Alternatively, we can estimate the time gap $T^\text{est,mom}$ from data
collected by a momentary detector, again assuming constant velocities of the
vehicles:
\begin{equation}
  T^\text{est,mom}_\alpha(t_\alpha) = \frac{s_\alpha(t_\alpha)}{v_{\alpha-1}(t_\alpha)}.
  \label{eq:timegap_mom}
\end{equation}

\subsection*{Data Preparation}
In total, we have investigated 184,171 datapoints in the Prototype dataset and
722,904 in the two other datasets.  Datapoints that were too close to the
downstream boundary needed to be discarded since no spatial leader could be
identified.  Furthermore, we ignored datapoints that were closer than
$\unit[3]{s}$ to a lane-changing event, leaving us with 146,213 datapoints from
the Prototype dataset and 675,660 from the I-80 and US-101 datasets.

Due to tracking or vehicle dimension detection errors, some spatial
and time gaps are negative or very small.  A small spatial gap
$s_\alpha$ leads to a very large inverse time-to-collision
$\tau_\alpha$ (see below) which would dominate any higher-order
moments of the $\tau_\alpha$ distribution.  Thus, we filtered the data
such that $s_\alpha\geq\unit[1]{m}$ and $T_\alpha=\geq\unit[0.1]{s}$
holds for every datapoint.  This filter removed further 3,070
datapoints (2.1\%) from our Prototype dataset extract and 11,755
(1.7\%) from the extract of the two later datasets.

\subsection*{Microscopic Fundamental Diagram and Stopped Traffic}

\begin{figure}
  \centering
  \begin{tabular}{cc}
  \includegraphics[width=.45\linewidth]{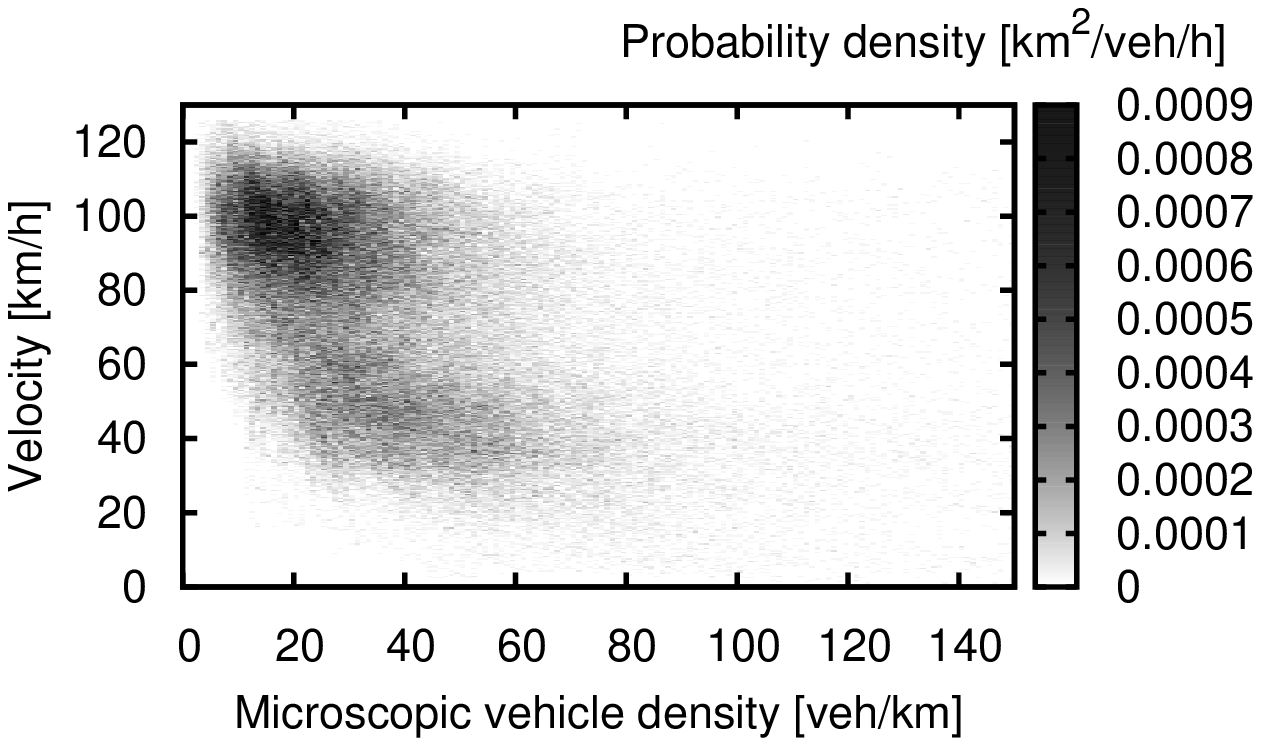} &
  \includegraphics[width=.45\linewidth]{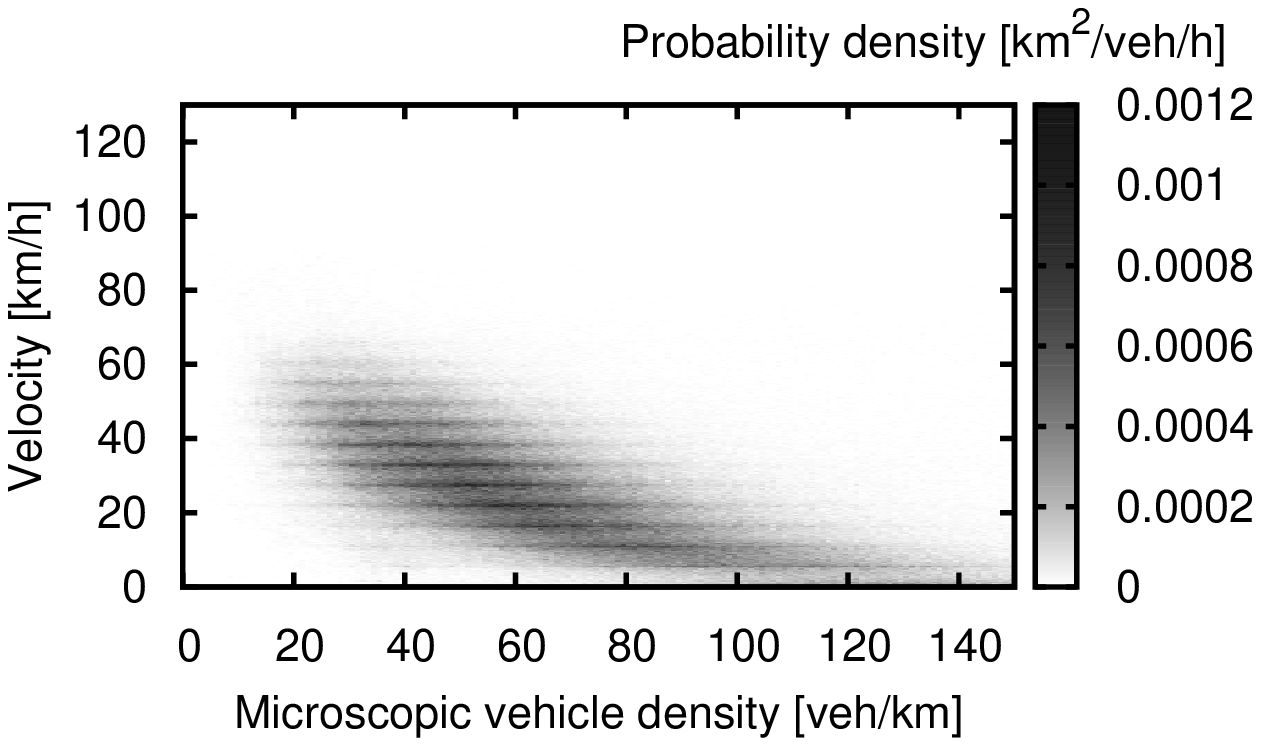}\\[-2ex]
  \includegraphics[width=.45\linewidth]{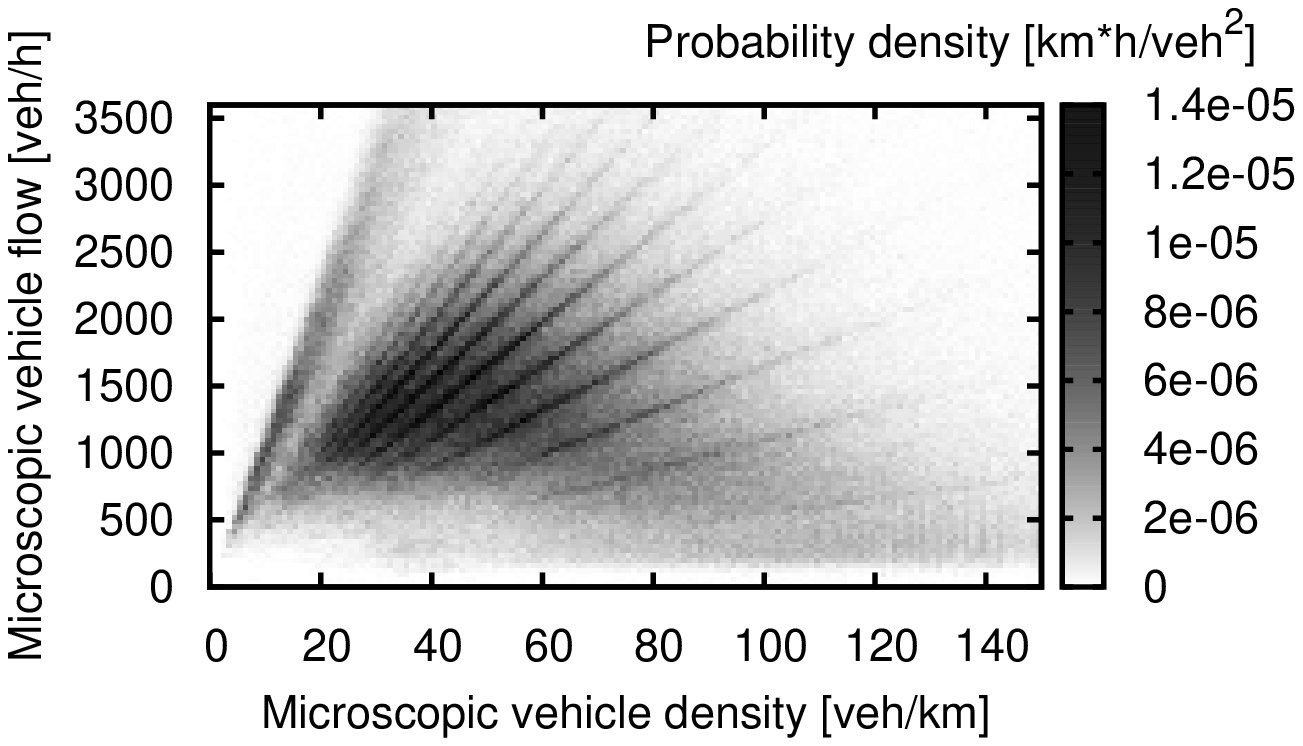} &
  \includegraphics[width=.45\linewidth]{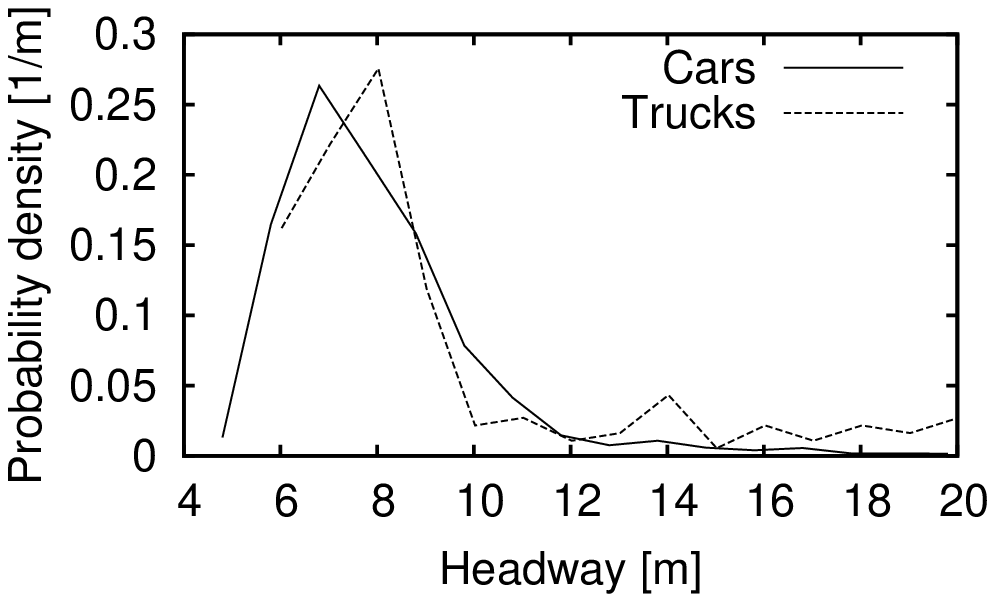} 
  \end{tabular}

  \caption{\label{fig:microfunddia}Probability density of the two-dimensional
  distribution (i.e., a ``two-dimensional histogram'') of the
  microscopic density $(x_{\alpha-1} - x_\alpha)^{-1}$ vs.\ velocity
  $v_\alpha$ in the Prototype dataset (upper left) and the two later
  I-80 and US-101 datasets (upper right).  In the bottom left plot,
  the probability density of the microscopic density vs.\ microscopic
  flow $T_\alpha^{-1}$ is shown.  The bottom right figure shows the
  probability density of the distribution of headways $x_{\alpha-1} -
  x_\alpha$ in stopped traffic ($v_\alpha < \unit[0.05]{m/s}$).  The
  mean value is $\unit[8.3]{m}$ for cars and $\unit[9.9]{m}$ for
  trucks.} 
\end{figure}

From the spatiotemporal measurements described above we can derive the inverse
of the space headway, $(\Delta x_\alpha)^{-1} = (x_{\alpha-1} -
x_\alpha)^{-1}$, and the inverse of the time headway, $(\Delta
t_\alpha)^{-1}$.  These quantities are more intuitively described as
``microscopic density'' and ``microscopic flow'', respectively, and will be
referred to by these names throughout this section.  For the Prototype dataset
and the combined other two datasets, we plotted the distribution of velocity
and microscopic density in Fig.~\ref{fig:microfunddia} (top row).  One clearly
sees that the Prototype dataset mainly features free traffic and some bound
traffic while the two later datasets feature only bound and jammed traffic.
Plotting microscopic flow vs.\ microscopic density for all three data sets, we
obtain the fundamental diagram (Fig.~\ref{fig:microfunddia}, bottom left).
The free flow part of the diagram is completely provided by the Prototype
dataset and the bound and jammed part is almost completely provided by the
I-80 and US-101 datasets. Notice that, in contrast to the Prototype dataset,
the later two sets exhibit stripes corresponding to the ``preferred
velocities'' as seen in Fig.~\ref{fig:originaldata}, which are much more
prominent when applying the same procedure to the original, unsmoothed data.

From the rich amount of data in the jammed traffic regime it is also
possible to determine the average headway of standing vehicles.  We
extracted all datapoints with velocities $v_\alpha < \unit[0.05]{m/s}$
and plotted the distribution of $\Delta x_\alpha$ in
Fig.~\ref{fig:microfunddia} (bottom right).  The mode is at
approximately $\unit[7]{m}$ for cars and $\unit[8]{m}$ for trucks
(with a smaller second peak at $\unit[14]{m}$).  However, the
distribution is right-skewed, so that the mean values are a little
higher: $\unit[8.3]{m}$ for cars and $\unit[9.7]{m}$ for trucks. Note
that, for principal reasons, this distribution cannot be obtained from
stationary detector data.

\clearpage
\subsection*{Time Gap Distribution}

\begin{figure}[b!]
  \centering
  \begin{tabular}{cc}
  \includegraphics[width=.475\linewidth]{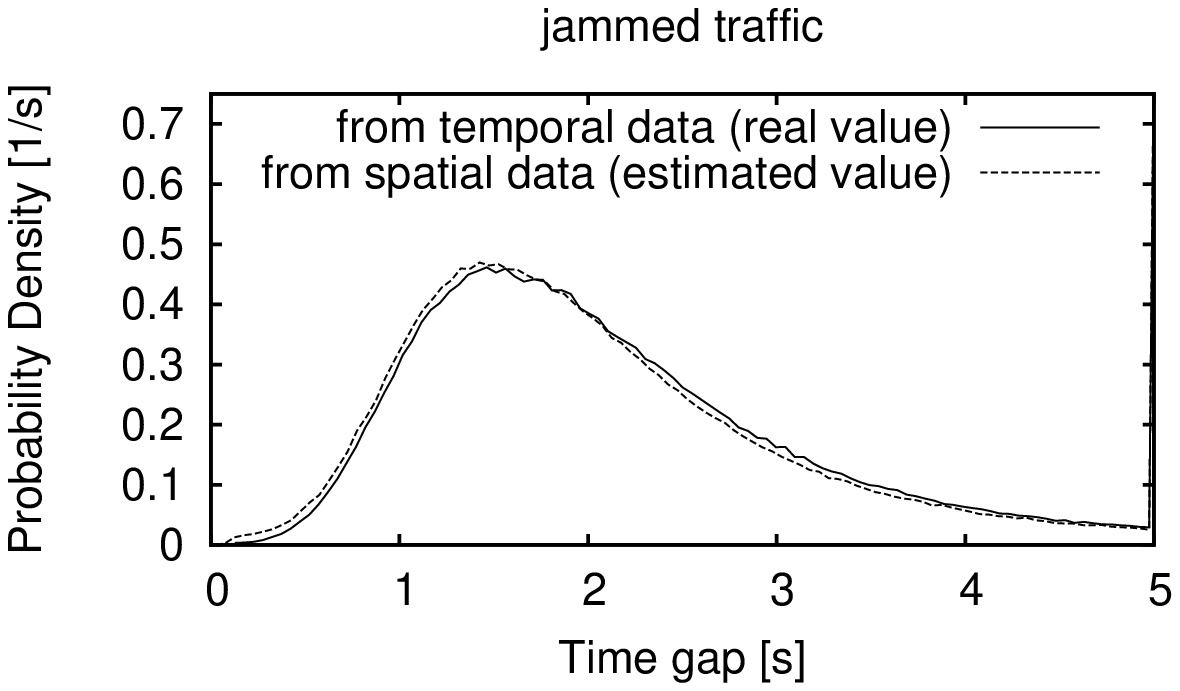} &
  \includegraphics[width=.475\linewidth]{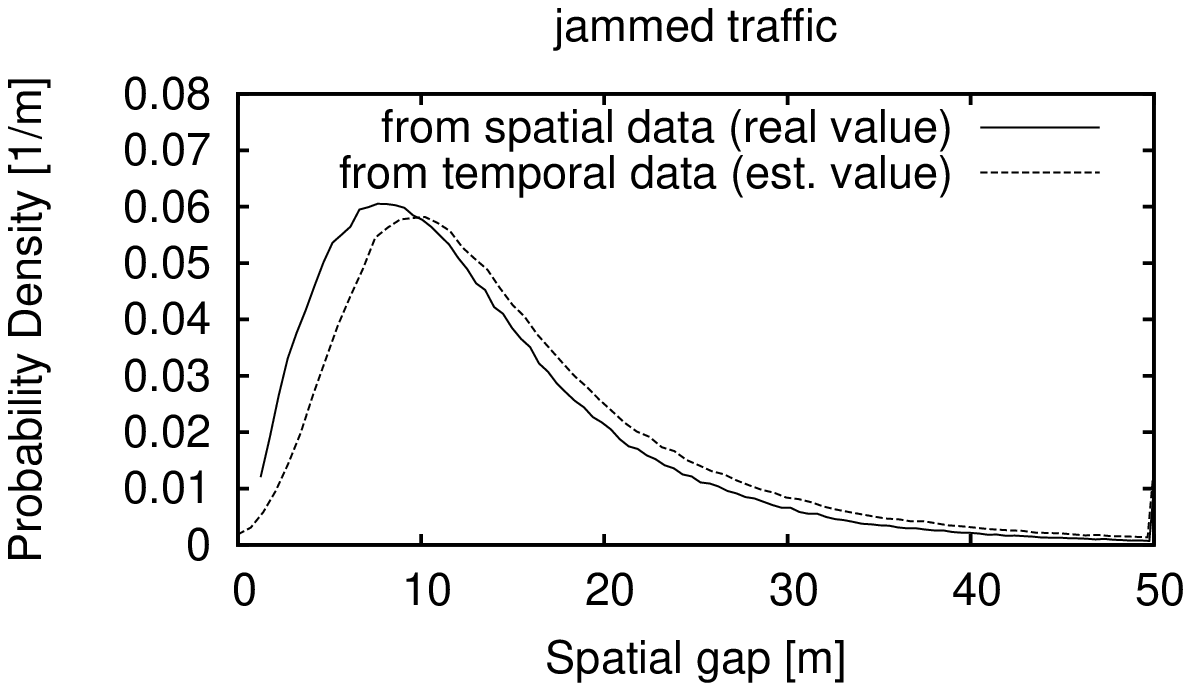}\\[-2ex]
  \includegraphics[width=.475\linewidth]{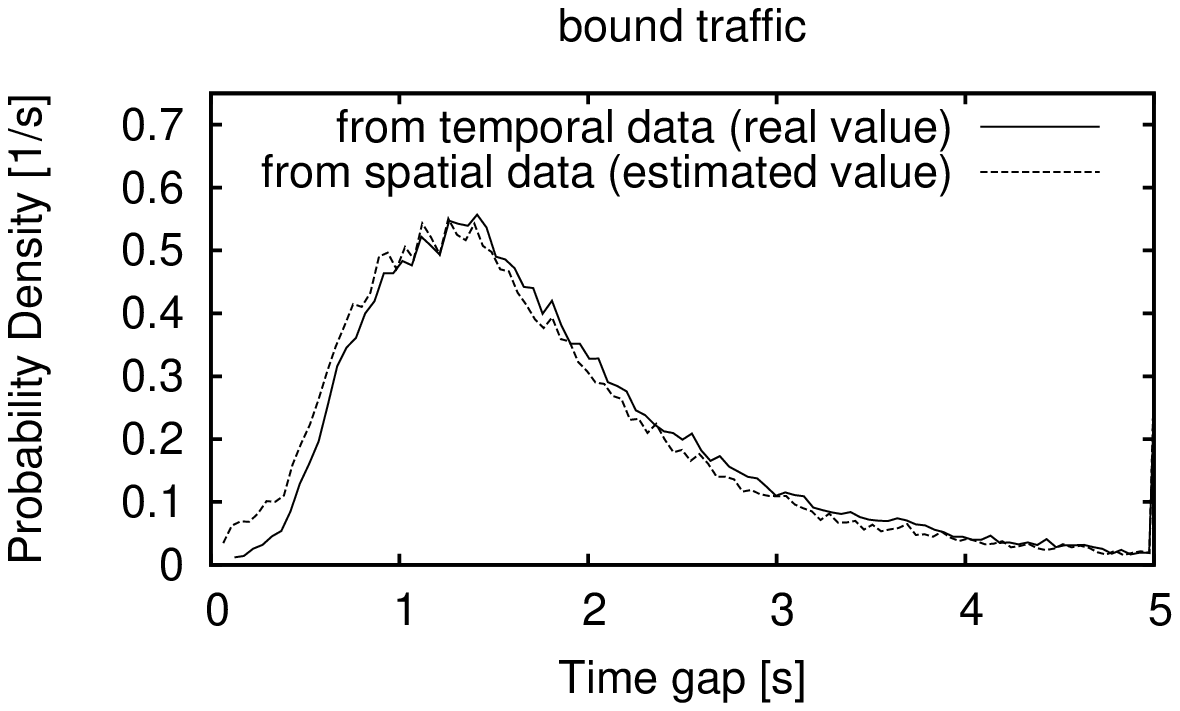} &
  \includegraphics[width=.475\linewidth]{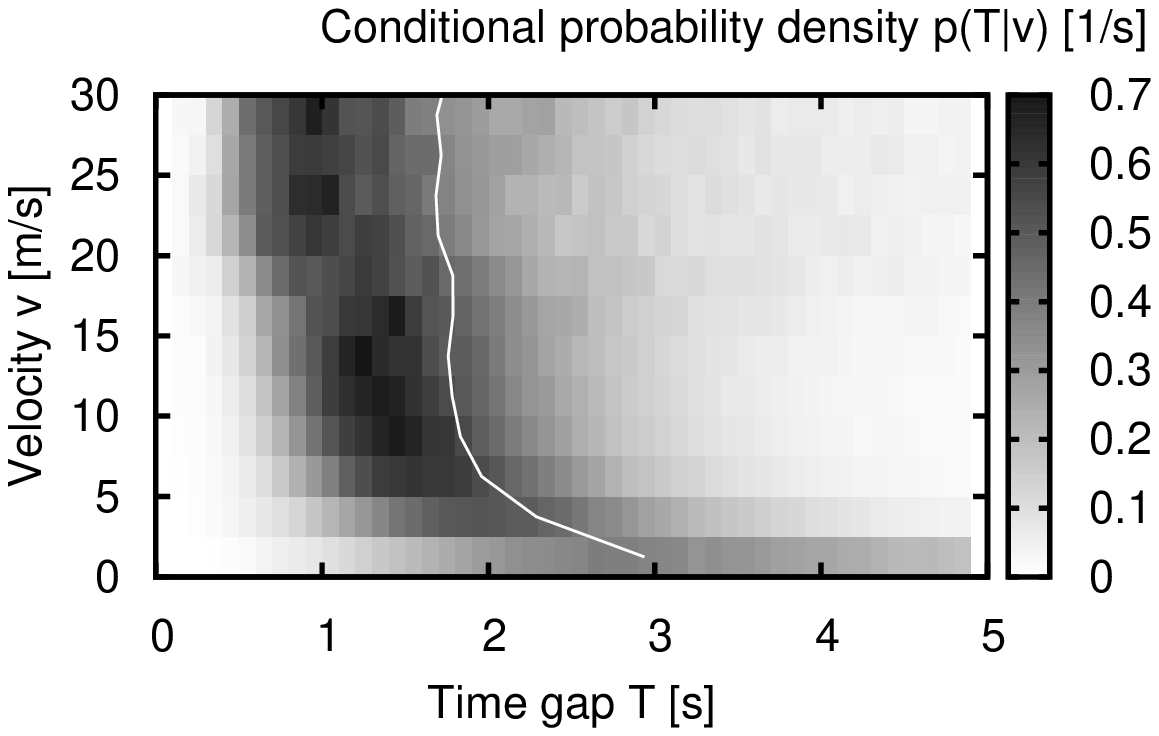}\\[-3ex]
  \includegraphics[width=.475\linewidth]{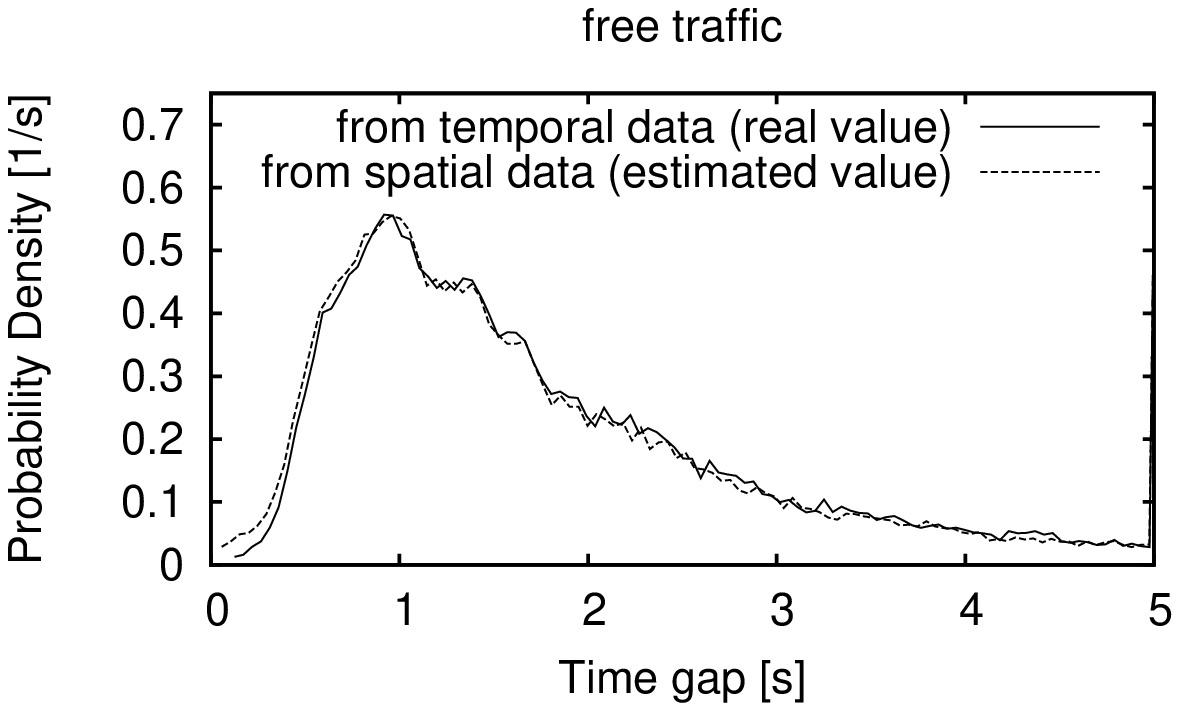} &
  \includegraphics[width=.475\linewidth]{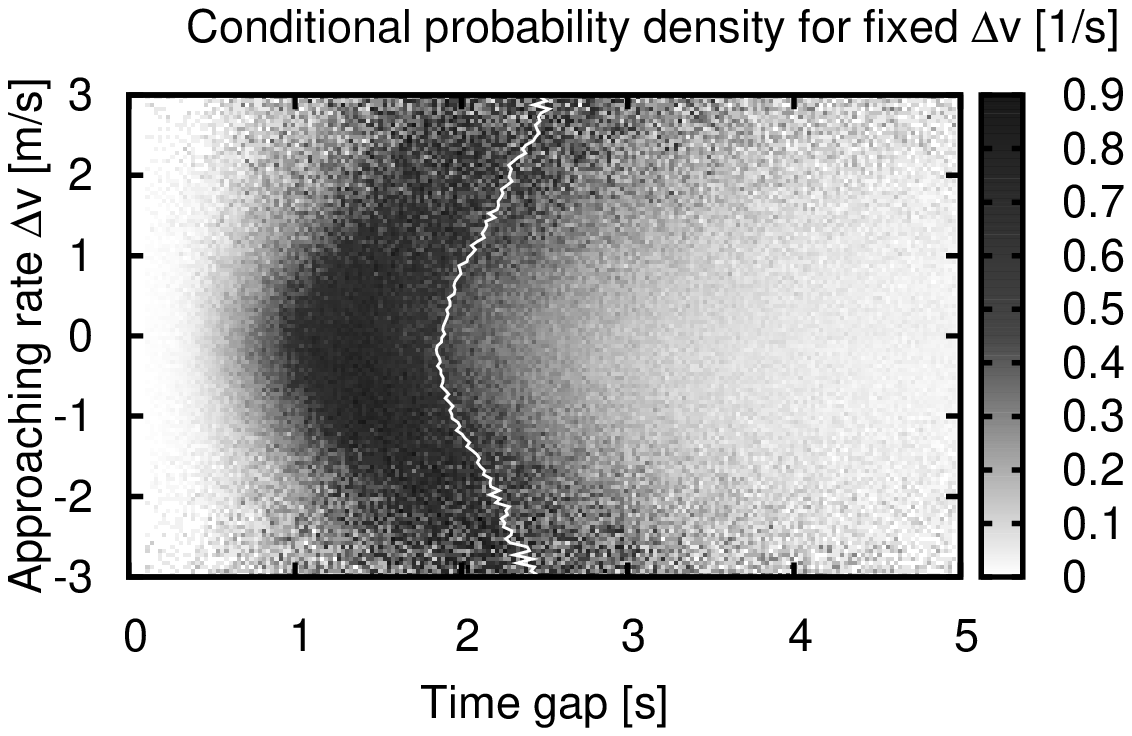}
  \end{tabular}

  \caption{The left column shows the distribution of time gap
  $T_\alpha$ in the different traffic regimes.  In the right column,
  we plotted the distribution of the spatial gap $s_\alpha$ in jammed
  traffic (top), the distribution of the time gap for different given
  velocities $v_\alpha$ (middle), and the distribution of the time gap
  for different given approaching rates $\Delta v_\alpha$ (bottom).
  The white lines show the mean value for each row of the plot, i.e.\
  the mean of the time gap for different values of $v$ or $\Delta v$.}
  \label{fig:timegap}
\end{figure}

Let us now look at the time gaps as defined in the 
Eqs.~\eqref{eq:timegap_pt} --~\eqref{eq:timegap_mom}.  In
Fig.~\ref{fig:timegap} we have plotted the time gap distribution in
three different traffic regimes: free traffic ($v >
\unit[22.2]{m/s}$), jammed traffic ($v < \unit[15]{m/s}$), and bound
traffic (intermediate velocities).  Furthermore, in every plot, the
real time gap $T_\alpha$ defined by Eq.~\eqref{eq:timegap} as obtained
from the trajectories is compared to the estimated time gap from
momentary measurement $T^\text{est,mom}_\alpha$ (cf.\
Eq.~\eqref{eq:timegap_mom}).  The first thing to note is the
remarkable indifference of the distributions to the measurement
method.  For comparison we have also plotted the spatial gap
distribution in jammed traffic (Fig.~\ref{fig:timegap}, top right),
which the stationary measurements shifts to larger values.  In the
other two traffic regimes the spatial gap distributions agree very
well.

Furthermore, it can be seen that the mode of the time gap distribution shifts
from approximately $\unit[1.5]{s}$ in jammed traffic to $\unit[1]{s}$ in free
traffic.  This effect is also visualized in the middle right plot of
Fig.~\ref{fig:timegap}.  The mean time gap is $\unit[2.6]{s}$ in jammed
traffic, $\unit[1.9]{s}$ in bound traffic, and $\unit[2.0]{s}$ in free
traffic.  In the bottom right plot we visualized another dependency of the
time gap: Although data becomes sparse towards larger values, there is a
significant tendency towards larger time gaps if the velocity difference
to the leading vehicle is large (regardless of whether approaching the vehicle
or falling behind).

Besides comparing time gaps measured by stationary detectors with time
gaps measured momentary detectors, there are also different ways to
determine the time gap with a stationary detector. The real time gap
is the time between the leader's rear bumper and the own front bumper
passing the detector~(Eq.~\eqref{eq:timegap}).  However, if detectors only produce passage
times and vehicle lengths and velocities, one needs to estimate the
timegap from the passage by assuming constant velocity of the leader
vehicle while passing the detector~(Eq.~\eqref{eq:timegap_pt}).  This error is very small in most
cases: only 10\% of our sample datapoints had an error in the estimate
from passage times $T^\text{est,pt}_\alpha$ that exceeded 10\% of the real time gap $T_\alpha$.

\clearpage
\subsection*{Time-to-Collision}
Another relevant quantity is the time-to-collision (TTC) which serves
as safety measure for traffic situations as it states the time left
until the vehicle will crash into its leader unless at least one of
the drivers changes speed \cite{Hirst,Minderhoud-TTC}. The TTC as a
spatial quantity is defined by Eqs.~\eqref{eq:spat_real_s}
and~\eqref{eq:spat_real_v} as
\begin{equation}\label{eq:ttc}
\tau_\alpha (t_\alpha) = \frac{s_\alpha(t_\alpha)}{\Delta v_\alpha(t_\alpha)}.
\end{equation}
The TTC can also be estimated from stationary (temporal)
measurements~\eqref{eq:spat_est_s} and~\eqref{eq:spat_est_v}:
\begin{equation}\label{eq:ttc_est}
\tau_\alpha^\text{est}(t_\alpha) = \frac{s_\alpha^\text{est}(t_\alpha)}{v_\alpha(t_\alpha) - v_{\alpha-1}(t_{\alpha-1}) }.
\end{equation}
We will now investigate the impact of the constant-velocity assumption
used to derive the TTC $\tau_\alpha^\text{est}$ from stationary
measurements. Since the TTC diverges for $\Delta v_\alpha=0$, it is
more convenient to discuss the TTC in terms of its inverse
$\tau_\alpha^{-1}=\Delta v_\alpha/s_\alpha$.

\begin{figure}
  \centering
  \includegraphics[width=.475\linewidth]{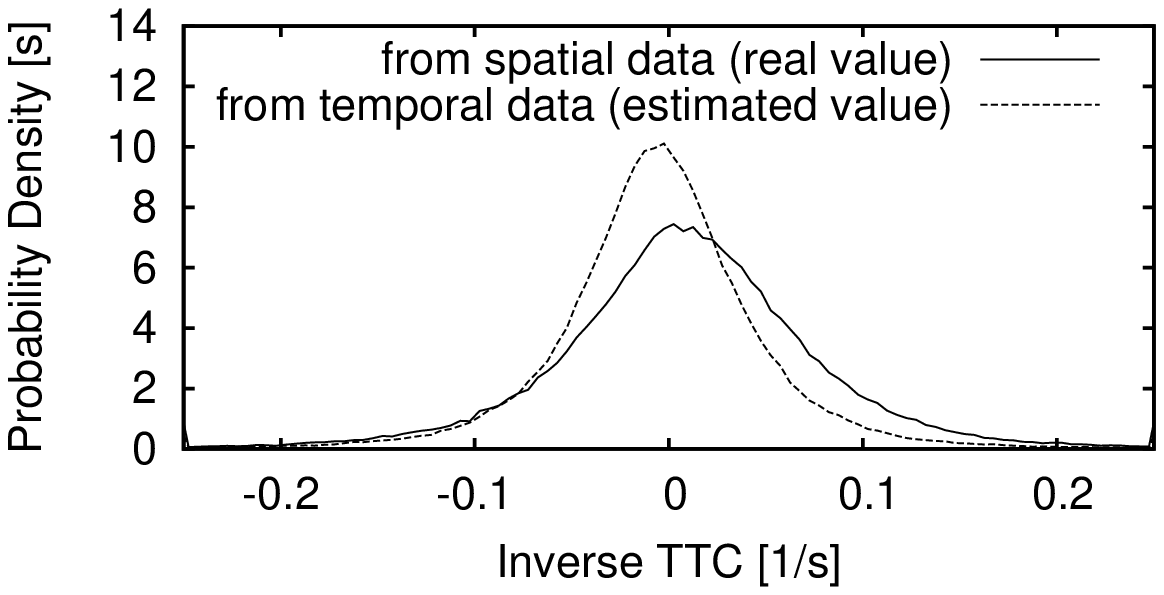}\hfill
  \includegraphics[width=.475\linewidth]{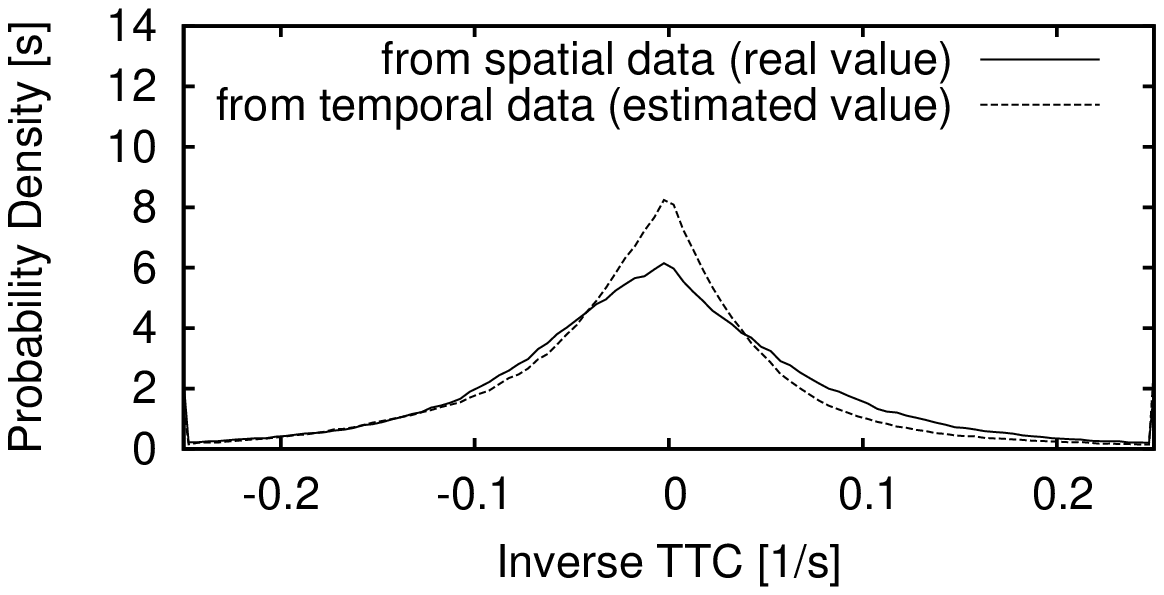}
  \includegraphics[width=.475\linewidth]{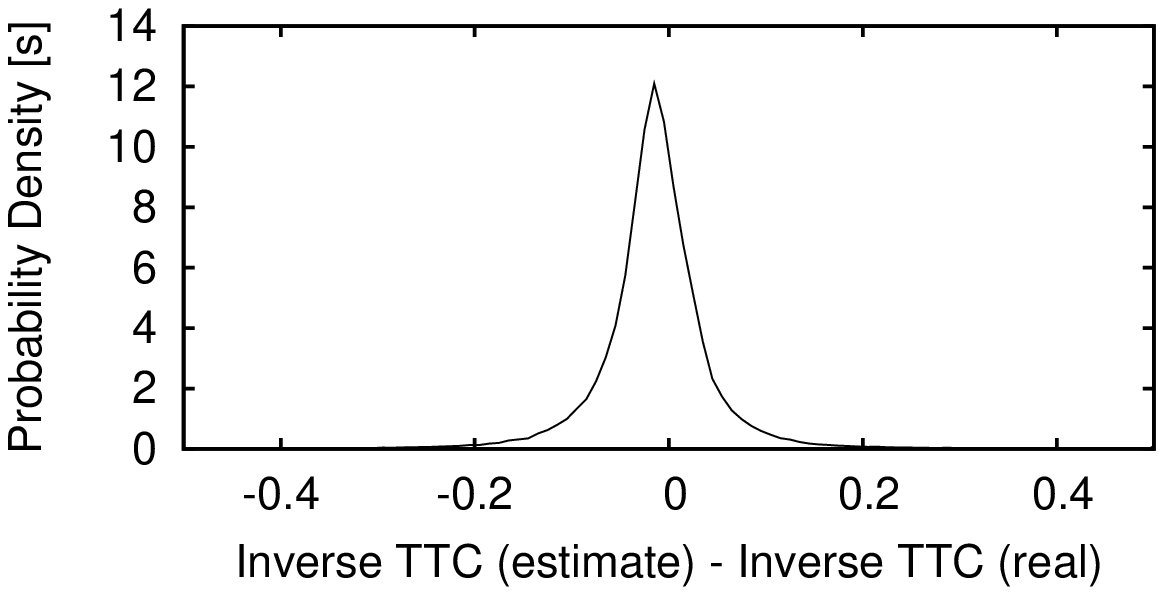}\hfill
  \includegraphics[width=.475\linewidth]{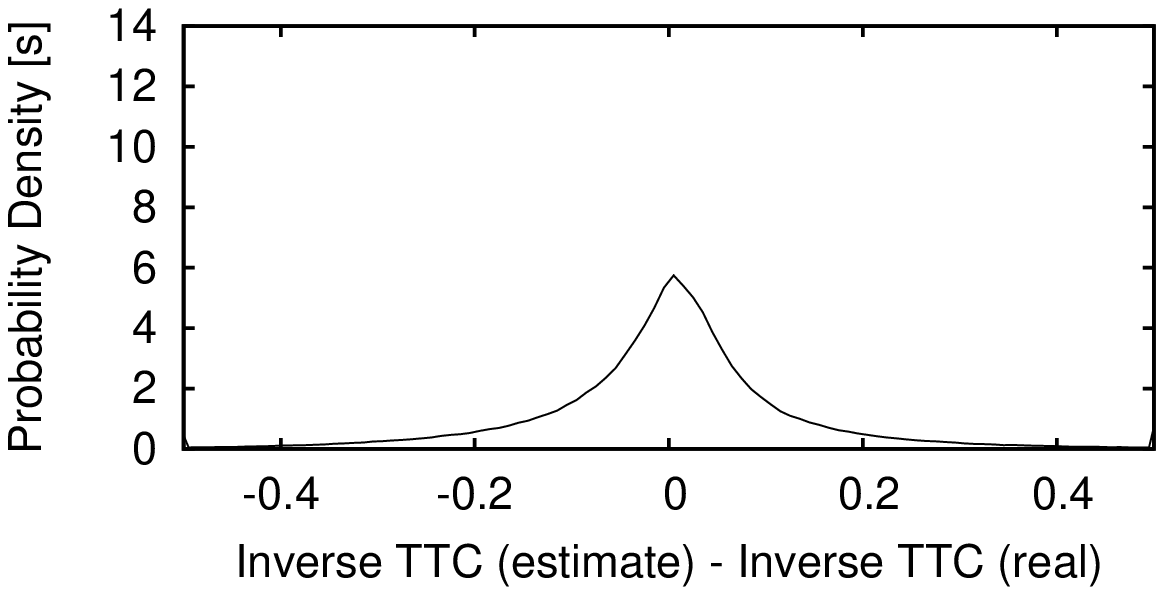}
  \caption{\label{fig:iTTC_dist_cmp}Distribution of the inverse
  time-to-collision $\tau_\alpha^{-1}$ in the Prototype dataset (left)
  and the two later datasets (right) compared to the estimated
  time-to-collision ${(\tau_\alpha^\text{est})}^{-1}$ obtained from
  stationary measurements.  The upper figures show both distributions
  while the lower figures show the distributions of the measurement
  errors $\Delta\tau_\alpha^{-1} = {(\tau_\alpha^\text{est})}^{-1} -
  \tau_\alpha^{-1}$.}
\end{figure}

In Fig.~\ref{fig:iTTC_dist_cmp} we plotted the distribution of the
inverse TTC in the Prototype dataset (left) and in the two later
datasets (right).  In contrast to the spatial and time gap
distributions, the inverse TTC distribution differs significantly
between the two measurement methods.  The inverse TTC is sensitive to
errors in the spatial gap, especially when the gap is small.
Therefore, we ignored inverse TTC values with absolute value larger
than $1$ when computing statistical properties of the distributions.
In this way, we ignored 0.59\% of all datapoints.

The mean of the absolute error $\Delta\tau_\alpha^{-1} :=
{ (\tau_\alpha^\text{est})}^{-1} - \tau_\alpha^{-1}$ is $0.00098$ in the
Prototype dataset and $-0.0134$ in the two later datasets.  The same
can be observed when splitting the data from all datasets into traffic
regimes as described above.  The mean error is $0.000045$ in jammed
traffic, $-0.0067$ in bound traffic, and $-0.0122$ in free traffic.
The variance of the errors is strongest in jammed traffic ($0.0236$),
while it is $0.00388$ in bound traffic, and $0.00225$ in free traffic.
Statistical properties of the inverse TTC distributions have been
collected into Table~\ref{tab:ittc_dist_statistics}.  One should
especially note that the skewness is consistently shifted towards
higher values by the stationary measurement.  This is visible in the
plots as well.

\begin{figure}
  \centering
  \includegraphics[width=.475\linewidth]{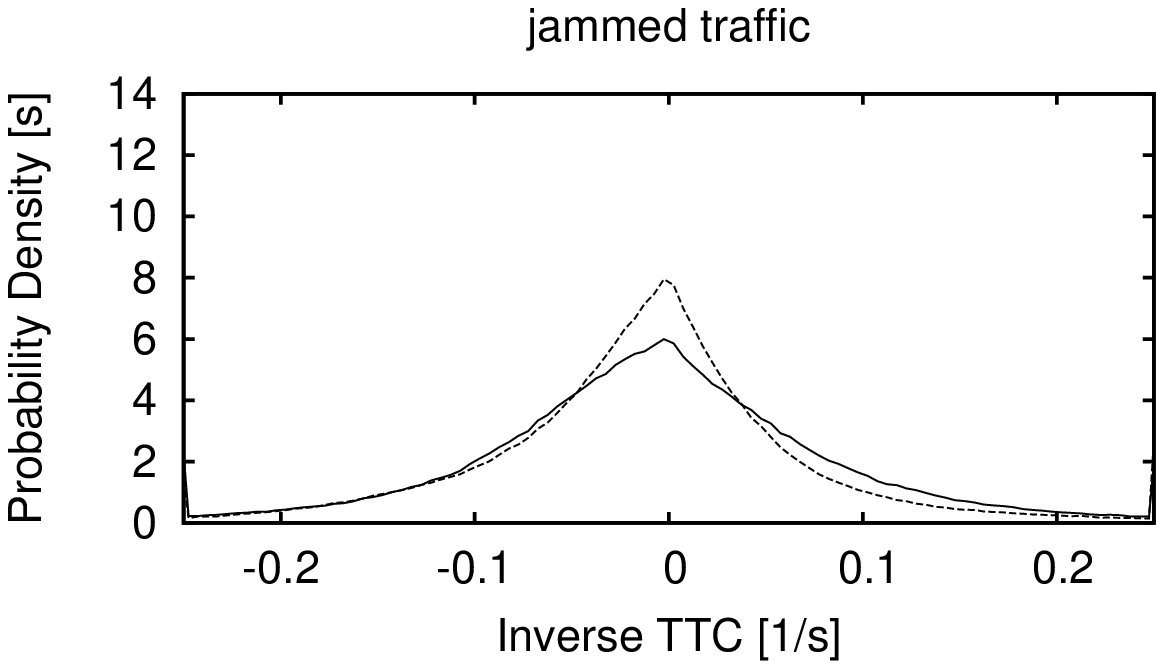}\hfill
  \includegraphics[width=.475\linewidth]{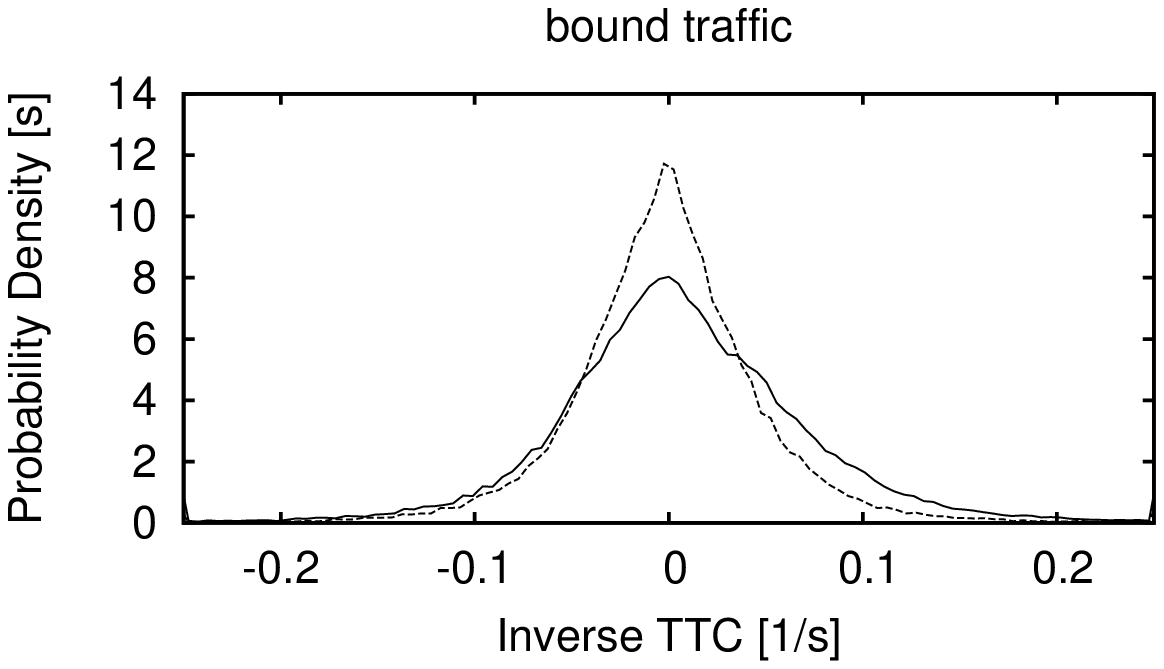}\\
  \includegraphics[width=\linewidth]{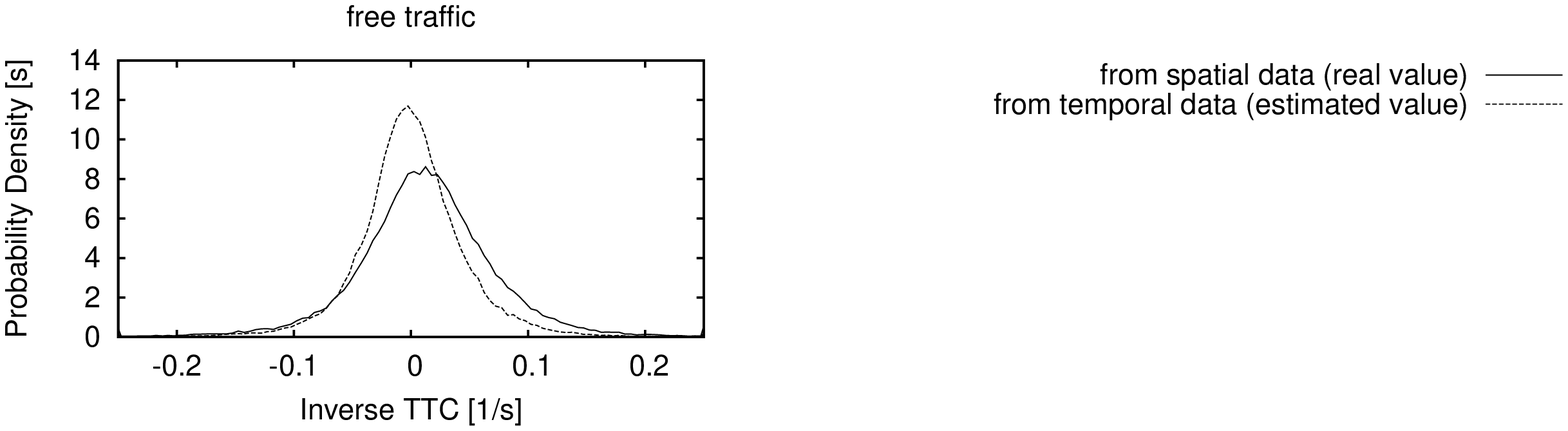}\\

  \caption{\label{fig:iTTC_dist_cmp_regimes}
  The figure shows
  distributions of the inverse time-to-collision $\tau_\alpha^{-1}$ of all data sets in
  different traffic phases compared to the estimated time-to-collision
  ${(\tau_\alpha^\text{est})}^{-1}$ obtained from stationary
  measurements.} 
 
\end{figure}

In view of the application of the TTC as safety measure, it is
particularly critical that stationary measurements consistently
decrease the probability of measuring a large positive
inverse-time-to-collision value which corresponds to a small positive
$\tau_\alpha$ indicating a dangerous traffic situation. For example in
free traffic (cf.\ Fig.~\ref{fig:iTTC_dist_cmp_regimes}), the fraction
of positive TTC values below $\unit[5]{s}$ (0.8\% of the datapoints)
which is considered as critical~\cite{Hirst,Minderhoud-TTC} is
underestimated by the stationary measurement by about a factor of
$2$. Thus, stationary measurements tend to euphemize the danger of
collision.


\begin{table}
  \centering
  \begin{tabular}{lp{2.5cm}p{2.5cm}p{2.5cm}p{2.5cm}}
    \toprule
    Dataset & Mean & Variance & Skewness & Sign change \\
    \midrule
    Prototype 
    & {\mbox{}\hfill 0.00874006\newline\mbox{}\hfill -0.00462994} 
    & {\mbox{}\hfill 0.00706569\newline\mbox{}\hfill 0.00422629}
    & {\mbox{}\hfill -0.264358\newline\mbox{}\hfill 0.16407}
    & {\mbox{}\hfill 19.2402\%\newline\mbox{}\hfill 7.63705\%} \\
    \midrule
    I-80/US-101 
    & {\mbox{}\hfill -0.00640645\newline\mbox{}\hfill -0.00542511}
    & {\mbox{}\hfill 0.0122408\newline\mbox{}\hfill 0.0120666}
    & {\mbox{}\hfill 0.0109132\newline\mbox{}\hfill 1.58624}
    & {\mbox{}\hfill 23.0413\%\newline\mbox{}\hfill 21.3802\%} \\
    \midrule
    jammed traffic 
    & {\mbox{}\hfill -0.00637853\newline\mbox{}\hfill -0.00633317}
    & {\mbox{}\hfill 0.0124101\newline\mbox{}\hfill 0.0121906}
    & {\mbox{}\hfill -0.000798933\newline\mbox{}\hfill 1.54999}
    & {\mbox{}\hfill 23.6459\%\newline\mbox{}\hfill 21.1149\%} \\
    \midrule
    bound traffic 
    & {\mbox{}\hfill 0.00624585\newline\mbox{}\hfill -0.000502914}
    & {\mbox{}\hfill 0.00699072\newline\mbox{}\hfill 0.00348709}
    & {\mbox{}\hfill -0.204996\newline\mbox{}\hfill 0.798759}
    & {\mbox{}\hfill 15.0853\%\newline\mbox{}\hfill 10.7964\%} \\
    \midrule
    free traffic 
    & {\mbox{}\hfill 0.0126101\newline\mbox{}\hfill 0.000389615}
    & {\mbox{}\hfill 0.00483055\newline\mbox{}\hfill 0.00260696}
    & {\mbox{}\hfill 0.179115\newline\mbox{}\hfill 0.459587}
    & {\mbox{}\hfill 16.6644\%\newline\mbox{}\hfill 5.58995\%} \\
    \bottomrule
  \end{tabular}
  \caption{Statistical properties of the inverse TTC distributions in the different
  datasets.  In the mean, variance, and skewness column, the top
  value is obtained from momentary measurements (the real value), while the
  bottom value is obtained from stationary measurements (the estimated value).
  In the sign change column, the top value states the amount of datapoints for
  which the stationary measurement determines a positive time-to-collision
  while the momentary measurement determines a negative value.  The bottom
  value gives the amount of datapoints for which the sign change is the other
  way round.  The jammed, bound, and free traffic dataset are combined from
  the Prototype and the two later NGSIM datasets.  A datapoint was assigned to
  jammed traffic if the vehicle's velocity was below $\unit[15]{m/s}$, to free
  traffic if $v_\alpha > \unit[22.2]{m/s}$, and to bound traffic, otherwise.}
  \label{tab:ittc_dist_statistics}
\end{table}

\clearpage
\subsection*{Lane Changes}

Besides the ability to compare stationary and momentary measurements, the
NGSIM trajectory data sets also provide a good basis to investigate lane
changes.  In order to determine the lane change duration, we collected all
lane changes in the NGSIM data.  However, from the processed video data
supplied with the NGSIM datasets it can be seen that sometimes the tracking
algorithm accidentally misplaced a vehicle across the lane boundary and back
after a few timesteps.  Also, sometimes drivers might have abort an already
begun lane change or quickly crossed two lanes.  Since we just want to look at
real and normal single-lane lane changes, we therefore filtered out all lane
changes that were closer than a certain threshold $\tau_\text{th}$ to another
lane change, which we chose to be $\tau_\text{th} = \unit[5]{s}$.  We also
sorted out lane changes that did not involve one of the four left-most lanes
in order to reduce the effect of the on-/off-ramp on our lane change analysis.

The former criterion was chosen to sort out
cases where drivers aborted an already begun lane change or where the
tracking algorithm accidently misplaced a vehicle across the lane
boundary.  The latter criterion ensures that we look at
discretionary lane changes only.

\begin{figure}
  \centering
  \includegraphics[width=.6\linewidth]{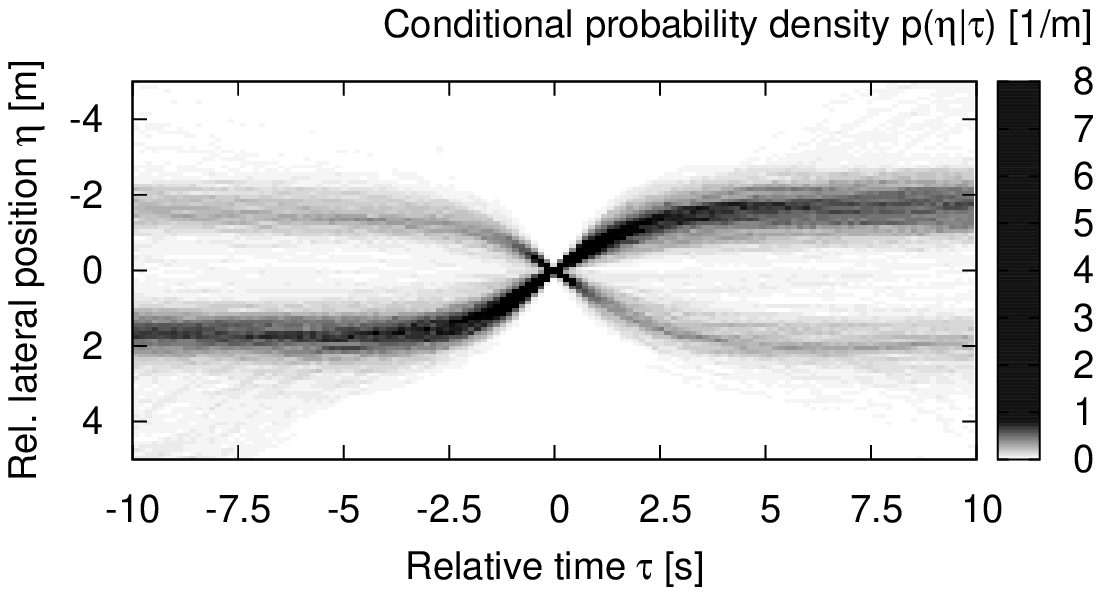}\\[-3ex]
  \includegraphics[width=.475\linewidth]{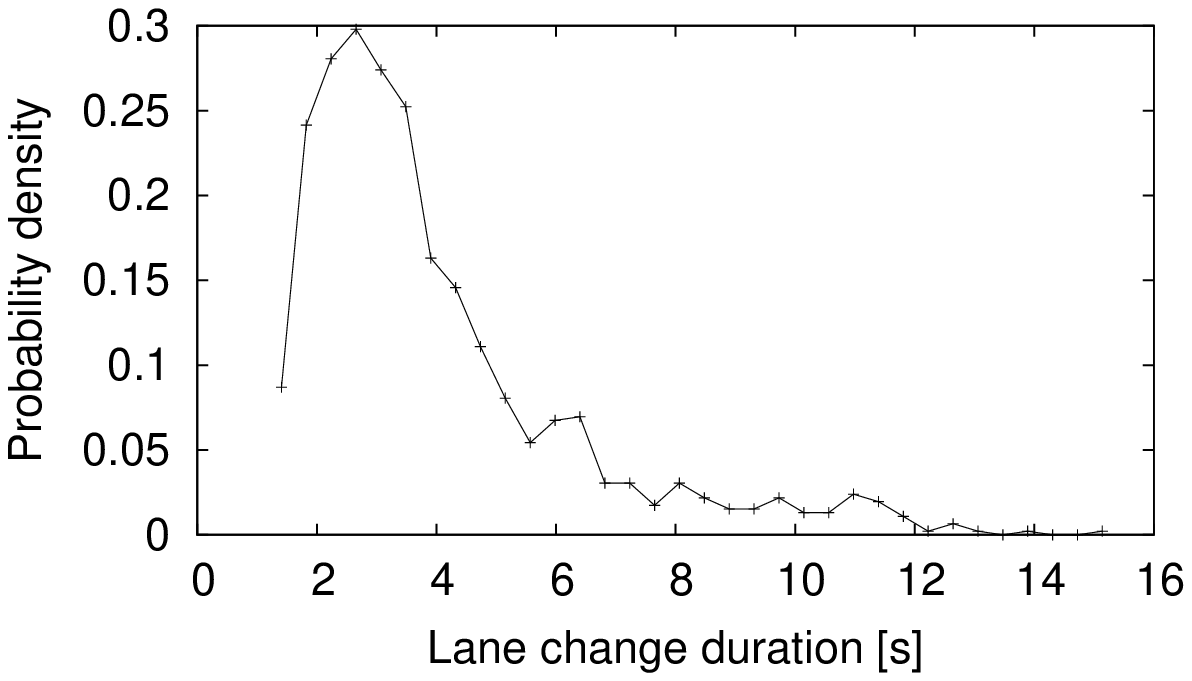}\hfill
  \includegraphics[width=.475\linewidth]{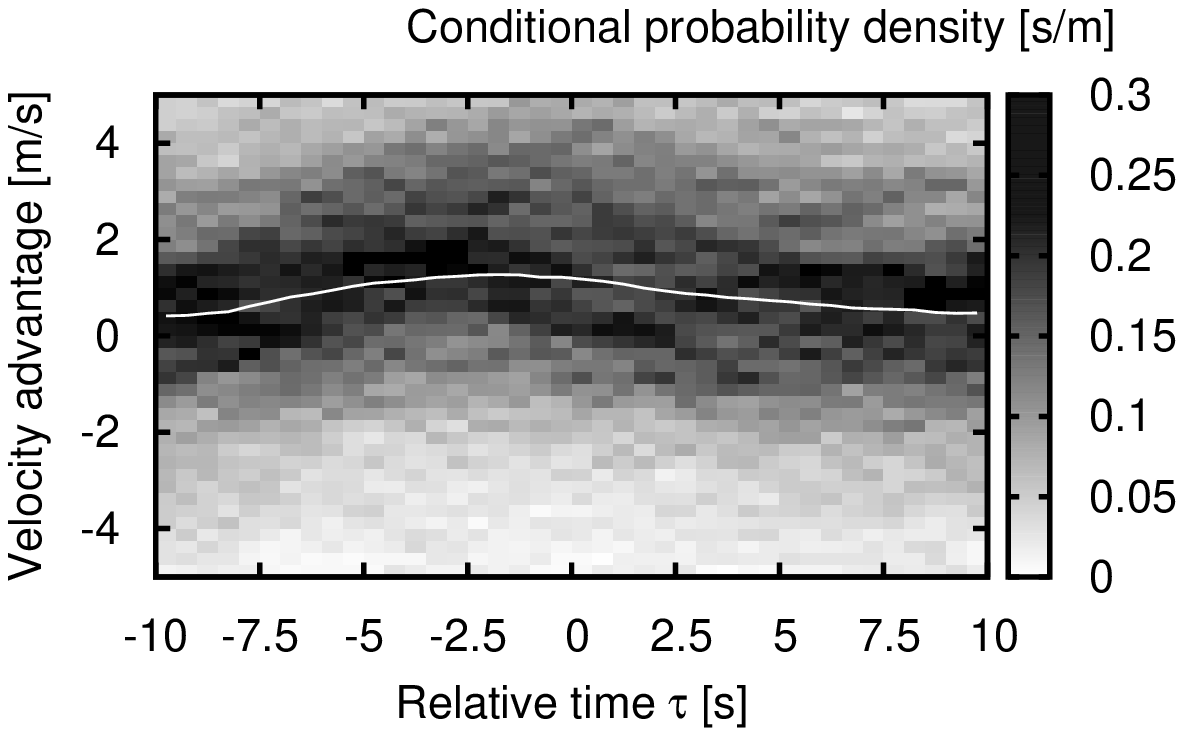}
  \caption{Lane changes: The upper plot shows the conditional
  probability $p(\eta|\tau)$ of finding a vehicle on lateral position
  $\eta$ relative to the lane boundary at a given time $\tau$ relative
  to the lane-changing event time.  The lower left plot shows the
  distribution of lane change durations $T_{\text{lc},\alpha}$
  according to the definition~\eqref{eq:lc_duration} given in the
  text.  The mean lane change duration is $\bar T_\text{lc} =
  \unit[(4.01\pm 2.31)]{s}$.  The lower right plot shows the
  conditional probability density of the velocity difference between
  the leader on the destination lane and the leader on the source lane
  for fixed times relative to the lane-changing event (the white line
  shows the mean value).}  \label{fig:lanechanges}
\end{figure}

With $\lambda_\alpha(t)$ denoting the lane used by vehicle $\alpha$ at time
$t$, a lane-changing event occurs at time $t_\text{lc}$ if
$\lambda_\alpha(t_\text{lc}) \neq \lambda_\alpha(t_\text{lc} + \Delta t)$
(where $\Delta t$ is the time interval between two consecutive datapoints of a
trajectory).  For each lane-changing event, we extracted a
20-second-environment of the trajectory with time, longitudinal and lateral
position relative to the lane-changing event:
\begin{align}
    \label{eq:lc_relative_time}
    \text{Relative time }\: & \tau := t - t_\text{lc}, \\
    \label{eq:lc_relative_poslong}
    \text{Relative longitudinal position }\: & \xi_\alpha(\tau) := x_\alpha(\tau + t_\text{lc}) - x_\alpha(t_\text{lc}). \\
    \label{eq:lc_relative_poslat}
    \text{Relative lateral position }\: & \eta_\alpha(\tau) := y_\alpha(\tau + t_\text{lc}) - y_\alpha(t_\text{lc}).
\end{align}
Then, we are able to produce a plot of the conditional probability
density $p(\eta|\tau)$ that a vehicle is at a relative lateral
position $\eta$ at a certain time $\tau$ relative to the lane-changing
event time (Fig.~\ref{fig:lanechanges}, top).  From this, we can
roughly estimate the lane change duration to approximately
$\unit[5-6]{s}$ by looking at the curvature of the two mode values
$\hat\eta_+(\tau) := \argmax_{\eta>0}\{ p(\eta|\tau) \}$ and
$\hat\eta_-(\tau) := \argmax_{\eta<0}\{ p(\eta|\tau) \}$.  This
procedure is similar to the approach done in
Ref.~\cite{toledo2007mdo}, where the lane change start and end time of
each trajectory were determined by looking at the curvature of the
lateral position $y_\alpha(t)$.  However, finding the correct point in
the curvature might be somewhat arbitrary, thus we will in the
following look at a more well-defined way to measure a lower bound of
the lane change duration.

The NGSIM vehicle detection algorithm does not only detect the vehicle
position but also its length $l_\alpha$ and width $w_\alpha$.  Since
the lane assignment algorithm works such that each datapoint is placed
into the lane where its mid-point front-bumper position $(x_\alpha,
y_\alpha)$ lies in, it is possible to determine the time where a
lane-changing vehicle first intruded the destination lane and the time
where it just completely left the source lane.  Given the
lane-changing event time $t_\text{lc}$ and the relative time and
position as defined in Eqs.~\eqref{eq:lc_relative_time}--\eqref{eq:lc_relative_poslat}, the
relative start time $\tau_\text{s}$ and end time $\tau_\text{e}$ of the
lane change may be defined as follows (a higher lane index
$\lambda_\alpha$ corresponds to a larger lateral position
$\eta_\alpha$):
\begin{align}
  \tau_\text{s} &= \begin{cases}
    \max\{ \tau  \mid \tau <0 \,\text{and}\, \eta_\alpha(\tau) + w_\alpha/2 < 0 \} &
    \text{if }\lambda_\alpha(t_\text{lc}) < \lambda_\alpha(t_\text{lc} + \Delta t), \\
    \max\{ \tau  \mid  \tau <0 \,\text{and}\, \eta_\alpha(\tau) - w_\alpha/2 > 0 \} &
    \text{otherwise}.
  \end{cases} \\
  \tau_\text{e} &= \begin{cases}
    \min\{ \tau \mid  \tau <0 \,\text{and}\,  \eta_\alpha(\tau) - w_\alpha/2 > 0 \} &
    \text{if }\lambda_\alpha(t_\text{lc}) < \lambda_\alpha(t_\text{lc} + \Delta t), \\
    \min\{ \tau \mid \tau <0 \,\text{and}\,   \eta_\alpha(\tau) + w_\alpha/2 < 0 \} &
    \text{otherwise}.
  \end{cases}
\end{align}
Then, of course, the lane change duration is obtained trivially from
\begin{equation}
  T_\text{lc} = \tau_\text{e} - \tau_\text{s}.
  \label{eq:lc_duration}
\end{equation}

In total, we have investigated 1231 lane changes, 1105 of which were
suitable to calculate $T_\text{lc}$ according to
Eq.~\eqref{eq:lc_duration}.  In the remaining 126 cases, either
$\tau_\text{s}$ or $\tau_\text{e}$ were undefined because the
corresponding condition was not fulfilled for any $\tau\in[-10,10]$
within the 20-second-environment around the lane change.  This can be
attributed to vehicle dimension detection errors or vehicle tracking
errors, both leading to a trajectory where the vehicle drives on the
lane boundary for some time.  Figure~\ref{fig:lanechanges} (bottom
left) shows the distribution of the lane change duration of the
examined lane changes.  One immediately notices that most lane changes
take somewhat about $\unit[3]{s}$ (mode value of the distribution), a
value already found valid for German highways back in
1978~\cite{sparmann1978saz}, which is, however, substantially
different from the one obtained by rule of thumb from the conditional
probability density $p(\eta|\tau)$.  The mean and standard variation
of the distribution are
\begin{equation}
  \bar T_\text{lc} = \unit[4.01\pm 2.31]{s}.
  \label{eq:NGSIM_Tlc}
\end{equation}
However, one should be aware that definition~\eqref{eq:lc_duration}
measures the time span where the vehicle occupies two lanes, which can
only be taken as a lower bound of the real lane change
duration. Including the preparation and possible post-processing of a
lane change, a value of $\unit[5-6]{s}$ might seem realistic. Since the
``real'' beginning of a lane change, the decision for making the lane change,
is impossible to measure, and the ``physical' beginning, the moment where the
driver starts to turn the wheel, is very difficult if not impossible to
measure, we think that our proposed definition is a good estimator for the
lane change duration, because it uses well-defined and easily measurable
quantities.

In the lower right, Fig.~\ref{fig:lanechanges} shows the conditional
probability density of the velocity difference between the leader on
the destination lane and the leader on the source lane for different
fixed times relative to the lane-changing event.  As indicated by the
white line, the mean value rises before the lane change by
approximately $\unit[1]{m/s}$.  This indicates that drivers perceive a
velocity advantage on the destination lane before performing the
lane-changing maneuver and take anticipationary actions.

%
%
\clearpage
\section*{Discussion and Future Research}
The availability of the NGSIM data sets spurred a considerable
research activity, particularly with respect to lane changing, where
larger-scale empirical investigations are now possible for the first
time. To date, most researchers only used the positional information
which allows, for example, to investigate the lane-changing rate, the
duration of lane changes, the gap-acceptance behavior, or the
propagation velocity of longitudinal density waves.

The full potential of the data, i.e., using the positional information
together with that for velocity and acceleration, has hardly been
tapped.  A possible reason is that the velocity and acceleration
information cannot be used directly since the noise of the positional
information is greatly increased by the necessary numerical
differentiations. In this paper, we have developed a filter to extract
more realistic velocity and acceleration information from the
positional data. Since the trajectories are comparatively short, we
included the boundary regions in the filtered output by reducing the
width of the necessary smoothing operations near the boundary.  This
implies determining the most efficient order of the smoothing and
differentiation operations of the filter since they do no longer
commute, and a wrong order may even lead to a systematic bias.

It must be noticed that it is inherently difficult to determine the
optimal filter parameters that eliminate most of the noise while
retaining the real information. This is particularly crucial for
mean-reverting quantities such as the accelerations, where large
smoothing time intervals will eventually suppress the whole
information. Clearly, further research is necessary to develop more
sophisticated, possibly nonlinear, filters.

The velocity and acceleration information of the trajectories can be
used in many ways. In this work, we investigate the systematic errors
in determining spatial quantities from temporal information, and vice
versa. The background is that spatial quantities such as the gap to
the leading vehicle, the density, or the times-to collision, are
usually estimated by single-vehicle data from stationary detectors,
i.e., by using temporal information.  Using ``virtual stationary
detectors'' that are fed with the trajectory data and simulating the
estimation procedure, we could quantitatively determine the resulting
estimation errors. Besides the well-known underestimation of the real
density of congested traffic, we found that the percentage of critical
values of times-to-collision is underestimated by a factor of 2 and
more when estimated from single-vehicle data. This clearly is relevant
for safety-related applications.

Another application field are empirical tests and parameter
calibrations for car-following and lane-changing models. In this work,
we showed that, prior to a discretionary lane change, there is a noisy
and small, but significant, velocity difference in favor of the
target lane. From this, we conclude that lane-changing decisions are
not only based on gaps and velocities, but also on velocity
differences, and possibly, on accelerations as considered in
Ref.~\cite{MOBIL-TRR07}.

More generally, the trajectory data allow, for the first time, to
empirically investigate the strategical and tactical actions for
preparing or facilitating a lane change~\cite{MOBIL-TRR07}. Apart from
the actions of the lane-changing driver, this also includes the
actions of the other drivers involved, such as cooperative actions of
the follower on the target lane to allow zip-like merging. This is
relevant for microscopic simulation software since it turned out to be
notoriously difficult to model realistic lane changes, particularly in
the case of mandatory changes in congested traffic.

The acceleration information of the data can also be used to
investigate to which extent the local traffic environment
(consisting, e.g., of the next-nearest and further leading vehicles)
influences the longitudinal driving behavior~\cite{HDM}. For example,
it has been proposed that the driving style is influenced by the local
velocity variance as determined from few leading vehicles~\cite{VDT}.

Finally, the velocity and acceleration information can be used to
determine the influence of traffic congestion on the fuel consumption
and emissions~\cite{Treiber-Fuel-TRB08}. Since reliable characteristic maps
are available for the instantaneous fuel consumption and emission
rates of various pollutants as a function of velocity and
acceleration, these quantities can now be estimated, for real
situations, with unprecedented accuracy.

\section*{Acknowledgments}
The authors would like to thank the Federal Highway Administration for 
providing the NGSIM trajectory data used in this study.


\begin{thebibliography}{10}

\bibitem{NGSIM}
{US Department of Transportation}, ``{NGSIM -- Next Generation Simulation},'',
  2007, {\tt http://www.ngsim.fhwa.dot.gov} -- Access date: May 5, 2007.

\bibitem{skabardonis2005eav}
A. Skabardonis, ``Estimating and Validating Models of Microscopic Driver
  Behavior with Video Data,'' Technical report, California Partners for
  Advanced Transit and Highways (PATH) (2005) .

\bibitem{lu2007fts}
X.-Y. Lu and A. Skabardonis, ``Freeway Traffic Shockwave Analysis: Exploring
  the NGSIM Trajectory Data,'' In {\em TRB 2007 Annual Meeting CD-ROM},
  (2007).

\bibitem{roess2007aof}
R.~P. Roess and J.~M. Ulerio, ``Analysis of Four Weaving Sections: Implications
  for Modeling,'' In {\em TRB 2007 Annual Meeting CD-ROM},   (2007).

\bibitem{zhang2007fga}
L. Zhang and V. Kovvali, ``Freeway Gap Acceptance Behaviors Based on Vehicle
  Trajectory Analysis,'' In {\em TRB 2007 Annual Meeting CD-ROM},   (2007).

\bibitem{goswami2007gab}
V. Goswami and G.~H. Bham, ``Gap Acceptance Behavior in Mandatory Lane Changes
  under Congested and Uncongested Traffic on a Multi-lane Freeway,'' In {\em
  TRB 2007 Annual Meeting CD-ROM},   (2007).

\bibitem{toledo2007mdo}
T. Toledo and D. Zohar, ``Modeling duration of lane changes,'' Transportation
  Research Record {\bf 1999,} 71--78 (2007).

\bibitem{choudhury2007mcl}
C.~F. Choudhury, M.~E. Ben-Akiva, T. Toledo, G. Lee, and A. Rao, ``Modeling
  Cooperative Lane Changing and Forced Merging Behavior,'' In {\em TRB 2007
  Annual Meeting CD-ROM},   (2007).

\bibitem{leclercq2007rpa}
L. Leclercq, N. Chiabaut, J. Laval, and C. Buisson, ``Relaxation phenomenon
  after lane changing: Experimental validation with NGSIM dataset,''
  Transportation Research Record {\bf 1999,} 79--85 (2007).

\bibitem{vu2007sow}
T. Vu, R. Roess, J. Ulerio, and E. Prassas, ``Simulation of a weaving
  section,'' In {\em TRB 2007 Annual Meeting CD-ROM},   (2007).

\bibitem{jin2007sof}
W.-L. Jin and L. Li, ``A study of first-in-first-out violation in freeway
  traffic,'' In {\em TRB 2007 Annual Meeting CD-ROM},   (2007).

\bibitem{webster2007tdl}
N.~A. Webster, T. Suzuki, E. Chung, and M. Kuwahara, ``Tactical Driver Lane
  Change Model Using Forward Search,'' In {\em TRB 2007 Annual Meeting CD-ROM},
    (2007).

\bibitem{alecsandru2007ard}
C. Alecsandru and S. Ishak, ``Accounting for random driving behavior and
  nonlinearity of backward wave speeds in the cell transmission model,'' In
  {\em TRB 2007 Annual Meeting CD-ROM},   (2007).

\bibitem{Kesting-Calibration-TRB08}
A. Kesting and M. Treiber, ``Calibrating Car-Following Models using Trajectory
  Data: Methodological Study,'' Transportation Research Record  (2008), in
  print.

\bibitem{Treiber-Fuel-TRB08}
M. Treiber, A. Kesting, and C. Thiemann, ``How Much do Traffic Congestion
  Increase Fuel Consumption and Emissions? Applying a Fuel Consumption Model to
  the NGSIM Trajectory Data,'' In {\em TRB Annual Meeting 2008 CD-ROM},
  (Transportation Research Board of the National Academies, Washington, D.C.,
  2008).

\bibitem{Hirst}
S. Hirst and R. Graham, ``The Format and Presentation of Collision Warnings,''
  in {\em Ergonomics and Safety of Intelligent Driver Interfaces}, Y. Noy, ed.,
  (Lawrence Erblaum Associates, New Jersey, 1997).

\bibitem{Minderhoud-TTC}
M.~M. Minderhoud and P.~H.~L. Bovy, ``Extended time-to-collision measures for
  road traffic safety assessment,'' Accident Analysis \& Prevention {\bf 33,}
  89--97 (2001).

\bibitem{sparmann1978saz}
U. Sparmann, ``Spurwechselvorg{\"a}nge auf zweispurigen
  {BAB-Richtungsfahrbahnen},'' Forschung Stra{\ss}enbau und
  Stra{\ss}enverkehrstechnik 263 (1978).

\bibitem{MOBIL-TRR07}
A. Kesting, M. Treiber, and D. Helbing, ``{General Lane-Changing Model MOBIL
  for Car-Following Models},'' Transportation Research Record {\bf 1999,}
  86--94 (2007).

\bibitem{HDM}
M. Treiber, A. Kesting, and D. Helbing, ``Delays, inaccuracies and anticipation
  in microscopic traffic models,'' Physica A {\bf 360,} 71--88 (2006).

\bibitem{VDT}
M. Treiber, A. Kesting, and D. Helbing, ``Understanding widely scattered
  traffic flows, the capacity drop, and platoons as effects of variance-driven
  time gaps,'' Physical Review E {\bf 74,} 016123 (2006).

\end{thebibliography}

\end{document}